# Ultrasensitive, Ultrafast and Gate-Tunable Two-Dimensional Photodetectors in Ternary Rhombohedral ZnIn₂S₄ for Optical Neural Networks

Weili Zhen[ab§], Xi Zhou[de§], Shirui Weng[a], Wenka Zhu[a*], and Changjin Zhang[ac†]

[a]High Magnetic Field Laboratory, Chinese Academy of Sciences, Hefei 230031, China

[b]University of Science and Technology of China, Hefei 230026, China

[c]Institutes of Physical Science and Information Technology, Anhui University, Hefei 230601, China

[d]The Interdisciplinary Research Center, Shanghai Advanced Research Institute, Chinese Academy of Sciences, Shanghai 201210, China

[e]School of Microelectronics, University of Chinese Academy of Sciences, Beijing 100049, China

*Email: wkzhu@hmfl.ac.cn

†Email: zhangcj@hmfl.ac.cn

§W.Z. and X.Z. contributed equally to this paper

**ABSTRACT**: The demand for high-performance semiconductors in electronics and optoelectronics has prompted the expansion of low-dimensional materials research to ternary compounds. However, photodetectors based on 2D ternary materials usually suffer from large dark currents and slow response, which means increased power consumption and reduced performance. Here we report a systematic study of the optoelectronic properties of well-characterized rhombohedral ZnIn₂S₄ (R-ZIS) nanosheets which exhibit an extremely low dark current (7 pA at 5

V bias). The superior performance represented by a series of parameters surpasses most 2D counterparts. The ultrahigh specific detectivity ($1.8 \times 10^{14}$ Jones), comparably short response time ($\tau_{rise}$ = 222 µs, $\tau_{decay}$ = 158 µs) and compatibility with high-frequency operation (1000 Hz) are particularly prominent. Moreover, a gate-tunable characteristic is observed, which is attributed to photogating and improves the photoresponse by two orders of magnitude. Gating technique can effectively modulate the photocurrent-generation mechanism from photoconductive effect to dominant photogating. The combination of ultrahigh sensitivity, ultrafast response and high gate tunability makes the R-ZIS phototransistor an ideal device for low-energy-consumption and high-frequency optoelectronic applications, which is further demonstrated by its excellent performance in optical neural networks and promising potential in optical deep learning and computing.

## INTRODUCTION

Two-dimensional (2D) materials are a class of atomically thin materials with extraordinary electrical, optical, physicochemical and magnetic properties.[1-4] In the last decade, 2D layered materials and their van-der-Waals (vdW) heterostructures have attracted considerable attention in the fields of transistors, chemical catalysis, energy storage/conversion and photodetectors.[3, 5-7] To date, hundreds of transistors and photodetectors based on 2D materials have been fabricated.[7] Early researches on 2D materials primarily focused on elemental or binary compound materials such as graphene, black phosphorus, boron nitride, $MoS_2$, $WS_2$ and InSe.[8-12] As a typical 2D elemental material, graphene has a high carrier mobility but no band gap, and black phosphorus is unstable in air. The carrier mobility of the most studied 2D binary material $MoS_2$ is not high enough. Therefore, intensive efforts are devoted to discovering new 2D materials to meet the demands of future electronics and optoelectronics. Recently, 2D ternary materials have received more and more attention for their rich physical properties that can be tuned by multiple degrees of freedom and give birth to new promising applications.[4] Representative 2D ternary materials include metal phosphorous trichalcogenides (MPT, e.g., $MnPS_3$, $CrPS_4$, $NiPS_3$),[2, 13, 14] ternary transition metal carbides (TMCs, e.g., $Ti_3AlC_2$ and $Mo_2GeC$),[15, 16] ternary transition metal dichalcogenides (TMDs, e.g., $Ta_2NiSe_5$, $Cr_2Ge_2Te_6$, $Fe_3GeTe_2$),[4, 17, 18] and some perovskites.[19-21] These systems not only exhibit outstanding physical, chemical and electronic properties, but also have other unique properties that are not available in 2D binary systems. It has been demonstrated that 2D ternary chalcogenides have a wide range of applications in electronics, optoelectronics, biosensors and catalysis.[22]

As a new member of 2D ternary chalcogenides, $ZnIn_2S_4$ (ZIS) is a layered wide bandgap semiconductor with a bandgap of 1.72−2.48 eV.[23] ZIS has three phases, namely cubic,[24] hexagonal[25] and rhombohedral structures.[26] ZIS with different crystal structures has different characteristics in semiconductor electronics,[23] spectroscopy[27] and chemical catalytic performance.[28] A lot of researches have been conducted on the hexagonal and cubic phases, such as photocatalysis under visible light,[29] electrochemical properties of lithium-intercalated compounds,[30] high-performance thermoelectricity,[31] etc. However, the rhombohedral ZIS (R-ZIS) is rarely studied. In particular, its low-dimensional physical and optoelectronic properties have not yet been explored.

In this work, we report the growth of high-quality R-ZIS single crystals by the chemical vapor transport (CVT) method and detailed in-depth characterizations. The photodetectors fabricated on the exfoliated R-ZIS flakes exhibit a broadband photoresponse from near UV to visible light. Unlike photodetectors based on other 2D ternary materials (e.g., $Ta_2NiSe_5$,[32] $Bi_2O_2Se$[33] and $ZrGeTe_4$[34]) that usually suffer from large dark currents meaning increased power consumption and reduced optoelectronic performance, the R-ZIS devices show an extremely low dark current (7 pA at 5 V bias) at room temperature. The superior performance represented by a series of parameters, i.e., specific detectivity ($1.8 \times 10^{14}$ Jones), responsivity (230 A $W^{-1}$), response time ($\tau_{rise}$ = 222 µs, $\tau_{decay}$ = 158 µs) and external quantum efficiency ($6.12 \times 10^4$%) for a 405 nm laser, surpasses most documented 2D counterparts. Moreover, a gate-tunable characteristic is observed, which is attributed to photogating and increases the responsivity by two orders of magnitude ($\sim 10^4$ A $W^{-1}$) for a back gate of 70 V. The combination of ultrahigh sensitivity, ultrafast response and high gate tunability makes the R-ZIS photodetector an ideal device for low-energy-consumption and high-frequency optoelectronic applications, which is further highlighted by its excellent performance in optical neural networks (ONN).

**EXPERIMENTAL SECTION**

**Sample growth.** R-ZIS single crystals were grown by the CVT method, using iodine as the transport agent (see Figure S1 in the Supporting Information). The molar ratio of Zn powder (99.99%, Alfa Aesar), In powder (99.99%, Alfa Aesar) and Se powder (99.99%, Alfa Aesar) was 1:2:4, and then the mixture was loaded in a silica tube which was flame-sealed in a vacuum of 1 Pa. Throughout the growth process, the content of iodine vapor was maintained at about 10 mg cm$^{-3}$. The silica tube was placed in a two-zone furnace, and the two zones were heated to 750 °C and 700 °C at a rate of 50 °C/h, respectively, and kept for 10 days. Then the furnace was shut down and cooled to room temperature. The size of the obtained crystals can reach $8 \times 8 \times 0.05$ mm$^3$.

**Characterization.** Single crystal and powder X-ray diffraction (XRD) measurements were performed on a Rigaku-TTR3 X-ray diffractometer using Cu Kα radiation. The cross section of the sample was processed by a focused ion beam (Helios Nanolab 600i). Transmission electron microscopy (TEM) and high-angle annular dark field scanning TEM (HAADF-STEM) images were obtained from a JEOL JEM-ARM300F TEM/STEM with a spherical aberration corrector. The X-ray photoelectron spectroscopy (XPS) survey spectra were recorded using an ESCALab MKII X-ray photoelectron spectrometer with an X-ray source of Mg Kα (1253.6 eV). A Renishaw In Via spectrometer was used to acquire temperature-dependent photoluminescence (PL) spectra, where the excitation pulse was generated by a 405 nm laser. Variable temperature Raman spectra were collected in a backscattering geometry on a Horiba JobinYvon T6400 spectrometer with a 532 nm laser. The atomic force microscopy (AFM) measurement was conducted on a Park XE7 system and the ultraviolet (UV)−visible spectra were measured on a UV−visible spectrometer (PerkinElmer LAMBDA 950).

**Device fabrication.** To fabricate R-ZIS based devices, R-ZIS flakes were first prepared by mechanical exfoliation using Nitto tape (SPV224) and transferred to a heavily p-doped Si substrate capped with 300 nm $SiO_2$. Then the lift-off photoresist (PMGI−SF6) was spin coated (5000 r/min) to cover the entire substrate, and baked on a hot plate at 200 °C for 5 min. Another photoresist (SPR955, 0.7) was spin coated on the substrate (2500 r/min) and baked at 100 °C for 90 s. Finally, laser direct write lithography (DaLI, miDALIX) was used to pattern the source/drain electrodes and Au (90 nm) (or Cr/Au, 10 nm/90 nm) was deposited on the substrate by thermal evaporation, followed by the standard lift-off process.

**Optoelectronic measurements.** The electrical measurements were performed on a probe station (Prmat) combined with a Keithley 2636B system source meter. The optoelectronic measurement setup is presented in Figure S11 in the Supporting Information.

## RESULTS AND DISCUSSION

Natural ZIS single crystals have three phases, namely cubic phase with space group $Fd\bar{3}m$ (No. 227, $a$ = 10.59 Å), hexagonal phase $P6_3mc$ (No. 186, $a$ = 3.85 Å, $c$ = 24.68 Å) and rhombohedral phase $R\bar{3}m$ (also known as trigonal polymorph, No. 166, $a$ = 3.86 Å, $c$ = 37.01 Å). The corresponding atomic ball and stick models are presented in Figure 1a–c. For the cubic spinel structure, zinc and indium atoms are coordinated by four and six sulfur atoms, respectively.[35] In the hexagonal phase, zinc and sulfur atoms form tetrahedral coordination, while half of the indium atoms have tetrahedral coordination with sulfur atoms, and the other half have octahedral coordination.[25] For the rhombohedral phase, zinc and indium atoms occupy tetrahedral coordination sites in a mixed manner.[26] The hexagonal structure is a thermodynamically stable phase, and the cubic structure is considered a high-pressure phase. The two can be transformed into each other under certain conditions.[36] The structures of hexagonal and rhombohedral phases are similar, both composed of sandwich layers (one layer of octahedra and two layers of tetrahedra). The interlayer couplings are weak vdW force. The main difference lies in the arrangement of sulfur and the occupancy of metal sites.

The R-ZIS single crystals with a yellow luster are characterized by single crystal XRD. As shown in Figure 1d, there are exclusively one group of peaks in the pattern that can be indexed as the (00*l*) diffractions of the rhombohedral phase (JCPDS PDF No. 00-049-1562), indicating that the sample is a pure single crystal with out-of-plane orientation along the *c* axis. The powder XRD pattern is also indexed, as shown in Figure S2 of the Supporting Information. The interplanar spacing in the [001] direction is calculated to be $37.03 \pm 0.03$ Å. Figure 1e shows the optical image of a typical R-ZIS nanosheet that is peeled from the bulk crystal. Another image showing more R-ZIS flakes is included in the Supporting Information (Figure S3). Similar to other layered materials,[37] the layer-dependent optical contrast confirms that R-ZIS is a layered material. Figure 1f shows an AFM image of the flake circled in Figure 1e. The thickness is determined to be 11.6 nm, corresponding to three unit cells along the *c* axis. The composition and valence state of R-ZIS nanosheets are further characterized by XPS. As shown in Figure 1g, the two peaks at 1044.6 eV and 1021.5 eV are identified as Zn $2p_{1/2}$ and $2p_{3/2}$ branches, respectively, which are consistent with the +2 valence state of Zn ions. Taking into account the splitting caused by spin-orbit coupling (SOC), the peaks at 444.4 eV and 452.2 eV are determined to be the In $3d_{5/2}$ and In $3d_{3/2}$ branches of In$^{3+}$ ions (Figure 1h). In addition, the peaks at 161.3 eV and 162.5 eV confirm the $-2$ valence state of S atoms (Figure 1i). Throughout the survey spectrum, there is no detectable impurity signal except for the C and O peaks, again confirming the high quality of our crystals.

In order to further check the crystal structure and chemical composition on the atomic scale, thinner few-layer R-ZIS nanosheets are prepared by ultrasonication and examined using aberration-corrected STEM. The reliability of using TEM to identify atoms in thin-flake samples is discussed in the Supporting Information. Figure 2a shows the low-magnification STEM image of an R-ZIS flake taken in the HAADF mode. Figure 2b,c show the corresponding selected area electron diffraction (SAED) and high-resolution TEM (HRTEM), respectively. The SAED and HRTEM patterns are obtained with the incident beam along the [001] zone axis. The SAED pattern shows six-fold symmetric diffraction spots, which is consistent with the hexagonal structure of R-ZIS and again confirms the high crystal quality. The HRTEM is used to determine the crystal structure of R-ZIS, and the results indicate that R-ZIS has high crystallinity (Figure 2c). The distance between the atomic layers marked by the white lines is about 0.34 nm, which can be characterized as the [100] plane. Due to the co-occupation of zinc and indium atoms, only a few layers (< 3) can accurately reveal the structural information of R-ZIS. Figure 2d shows the low-magnification HAADF-STEM image of another R-ZIS nanosheet, where the symbols A and B denote few-layer and single-layer zones, respectively. As shown in the atomic-resolution HAADF-STEM image of zone B (Figure 2e), the 0.33 nm interplanar distance matches well with the $d_{100}$ and $d_{010}$ spacings, and the measured dihedral angle is exactly 60°, which is consistent with the theoretical value.[26] In addition, the enlarged image in Figure 2f clearly shows the presence of Zn vacancies. These vacancies are expected to significantly affect the spectroscopic characteristics of R-ZIS. The SEM and TEM studies performed on the cross section of R-ZIS prepared by focused ion beam are presented in Figure S4 in the Supporting Information. As shown in Figure 2g–i, the element mapping of the energy dispersive X-ray spectroscopy (EDX) reveals the uniform distribution of Zn, In, and S elements. The ratio of Zn, In and S is 12.47:31.55:55.98, which is close to the stoichiometric ratio 1:2:4. The spectroscopy with more detailed information is shown in Figure S5 in the Supporting Information. Note that the atomic ratio reveals a certain number of Zn vacancies, which is consistent with the observation through the atomic-resolution HAADF-STEM.

For the characterization of 2D layered materials, Raman spectroscopy is usually used because it contains rich information about crystal structure and phonon vibrations.[38] The layer-dependent Raman spectra can show significant changes in the intensity and position of phonon vibrations as a result of the change of vdW force between layers.[39] Figure 3a shows the Raman spectra of R-ZIS of various thicknesses from bulk to three layers taken at room temperature. The $R\bar{3}m$ space group contains 21 vibrational modes in the center of the Brillouin zone, which are described by the irreducible representations of the $D_{3d}$ point group.[40] Among these vibrational modes, the $A_{1g}$ and $E_g$ modes are Raman active. For bulk R-ZIS, four peaks are detected in the range of $200-450$ cm$^{-1}$, i.e., 248 cm$^{-1}$, 310 cm$^{-1}$, 356 cm$^{-1}$ and 364 cm$^{-1}$, which are labeled P1, P2, P3 and P4, respectively. This is in consistent with previous reports.[40] The P1 and P2 peaks are the $A_{1g}$-like vibrations corresponding to the out-of-plane breathing modes, and the P3 and P4 peaks show the $E_g$-like features corresponding to the in-plane breathing modes. As the thickness decreases, the P2 peak shows an obvious red shift, which agrees with the decrease of the vdW force in the thinner samples.

In order to investigate the interlayer thermal expansion and atomic vibration of the 2D R-ZIS flake, the 44 nm thick sample is further characterized by temperature-dependent Raman spectroscopy in the temperature range of 78–300 K. As seen in Figure 3b, all four peaks show a blue shift as the temperature decreases, which is the result of the phonon hardening process.[41] Figure 3c shows the Raman peak positions as a function of temperature, which can be fitted by the linear equation $\omega(T) = \omega_0 + \chi T$, where $\omega_0$ represents the Raman peak position at 0 K and $\chi$ is the first-order temperature coefficient. This method is usually used to describe the temperature dependence of the Raman shift of nanomaterials.[42, 43] For P1, P2, P3 and P4, the resulting $\chi$ is −0.01, −0.026, −0.013 and −0.024 cm$^{-1}$ K$^{-1}$, respectively. It is found that the P2 mode is more sensitive to temperature, which may be related to the vdW force between layers. These $\chi$ values are greater than those of MoS$_2$,[44] graphene[45] and SnSe$_2$,[46] indicating that R-ZIS flakes possess relatively strong interlayer interactions.

Absorption and PL spectroscopy are further performed to study the electronic structure of R-ZIS. Figure 4a shows the UV−visible absorption spectrum taken on a bulk crystal. The crystal has strong light absorption in a wide range of 300 nm to 600 nm, which is consistent with previous studies and can be attributed to intrinsic bandgap absorption, i.e., the result of electron hopping from the valence band to the conduction band. The broadened absorption edge with a long tail may originate from transitions within the bandgap,[47] due to the presence of sub-bands caused by Zn and S vacancies. The optical bandgap $E_{gap}$ can be calculated by the Kubelka-Munk equation $\alpha h v = A(h v - E_{gap})^{n/2}$,[48] where $\alpha$ is the absorption coefficient, $h$ is Planck's constant, $v$ is the light frequency, and $A$ is a constant. The absorption coefficient can be well functionalized in the form of $(\alpha h v)^2$ above the bandgap, suggesting that bulk R-ZIS is a direct bandgap semiconductor. The $E_{gap}$ is estimated to be 2.04 eV, which is less than the value determined in a previous study, i.e., 2.4 eV.[29]

The reduction of $E_{\text{gap}}$ is attributed to the doping level induced by defects and the structural reason, namely the superposition of the molecular orbitals due to the co-occupancy of Zn and In.[26] Therefore, the influence of the doping level and unique crystal structure of R-ZIS cannot be simply ignored, which is expected to affect the performance of R-ZIS devices.

It is demonstrated that PL spectroscopy can also be used to examine the intrinsic optical properties and electronic structure as well as the influence of defects on the physical properties of materials.[49] Figure 4b shows the PL spectra of R-ZIS flakes of different thicknesses ranging from bulk to three layers. A broad emission in the range of 600 nm to 800 nm is observed in all samples. It can be identified as two peaks, i.e., the peak at 607 nm (marked with a light blue vertical bar) and the peak at the longer wavelength (labeled A). The peak at 607 nm corresponds to an energy gap of 2.04 eV and does not change with the flake thickness, which is attributed to the bandgap emission. Emission A shows a clear thickness dependence. For the bulk sample, the peak appears at 670 nm, which corresponds to a lower energy gap and is associated with the sub-band emission.[50] The sub-band or doping level should originate from the Zn and S vacancies. Figure 4d is a schematic diagram of energy levels, in which $E_{\text{gap}}$ is the bandgap and A is the transition between the doping level represented by the dashed line and the conduction band. Generally, defects in thick layered semiconductors can cause a quenching effect,[51] making it difficult to detect the bandgap emission. This explains why the bandgap emission is not as prominent as emission A. As the thickness decreases, the intensity of peak A becomes weaker and the position shifts to shorter wavelengths, namely from 670 nm in the bulk sample to 634 nm in the three-layer sample, suggesting an evolution of doping level with the thickness. It has been demonstrated that thick samples generally contain a higher defect concentration.[50] With the decrease of thickness, the reduced defect concentration lowers the doping level. The weakened PL intensity is due to the less luminescent material in the thinner samples. Moreover, the temperature dependence of PL spectroscopy is performed to study the influence of electron distribution on the excitation and recombination process. Figure 4c shows the PL spectra of the 44 nm thick R-ZIS nanosheet taken at different temperatures from 78 K to 293 K. It can be seen that the PL intensity decreases with the increase of temperature, which is linked to the stronger defect recombination at higher temperature.[50] No obvious peak shift is observed, indicating that the energy levels are stable at various temperatures. It is expected that these excitations will play an important role in optoelectronic devices such as photodetectors, especially at the nanometer scale.

In order to systematically study the photoresponse of R-ZIS, devices based on the exfoliated nanosheets have been fabricated. Figure 5a is a schematic illustration of the R-ZIS photodetector. The channel area of the device is about 20 μm (length) × 6 μm (width), and the thickness of the active material is 40 nm. Figure 5b shows the current-voltage ($I_{ds}-V_{ds}$) characteristics taken in the dark and under laser illumination of different wavelengths (i.e., $\lambda$ = 405 nm, 450 nm, 488 nm and 515 nm) tuned to a constant power of 5.3 mW. The symmetrical $I_{ds}-V_{ds}$ curves indicate relatively good contact between Au electrodes and R-ZIS. We have also fabricated devices with Cr electrodes as source and drain, but the $I_{ds}-V_{ds}$ curves are asymmetric at low powers (≤ 14 μW), suggesting a relatively large Schottky barrier between Cr and R-ZIS (Figure S6 in the Supporting Information). An ultralow dark current ($I_{dark}$) is always essential to the high performance and low energy consumption of photodetectors with high responsivity and sensitivity.[52] In our nanoscale R-ZIS photodetector, an extremely low $I_{ds}$ is observed, i.e., 7 pA at $V_{ds}$ = 5 V, corresponding to a current density of $J_{dark}$ = 2.92 × 10$^{-3}$ A cm$^{-2}$. This room-temperature $I_{dark}$ is outstanding compared with the reported 2D semiconductors (see Table S1 in the Supporting Information). Under illumination, the $I_{ds}$ increases rapidly with the generation of electron-hole pairs. The time-resolved photocurrent obtained at various wavelengths shows that the most significant response occurs at $\lambda$ = 405 nm (Figure S7 in the Supporting Information). The ON/OFF ratio is calculated as 7.5 × 10$^5$, 2.0 × 10$^5$, 2.5 × 10$^4$ and 5.3 × 10$^3$ for $\lambda$ = 405 nm, 450 nm, 488 nm and 515 nm, respectively.

To further evaluate the optoelectronic performance, parameters such as responsivity ($R = \frac{I_{ph}}{PS}$, where $I_{ph} = I - I_{dark}$, $P$ is the laser power density and $S$ is the illuminated area), specific detectivity ($D^* = R\sqrt{\frac{S}{2eI_{dark}}}$, where $e$ is the elementary charge), external quantum efficiency (EQE = $\frac{hc}{e\lambda}R \times$ 100%, where $c$ is the speed of light in vacuum) and photosensitivity linear dynamic range (LDR = $20\log\frac{I_{ph}}{I_{dark}}$) are calculated. The reason for using the dark current instead of total noise current to calculate $D^*$ is explained in the Supporting Information. Figure 5c,d show $R$, $D^*$ and EQE as a function of laser wavelength, respectively. The most prominent parameters are $R$ = 200 A W$^{-1}$, $D^*$ = 1.58 × 10$^{14}$ Jones and EQE = 6.12 × 10$^4$ % at $\lambda$ = 405 nm and $P$ = 212 μW mm$^{-2}$. The responsivity is superior to commercial silicon-based near-UV detectors (0.1−0.2 A W$^{-1}$ for $\lambda$ < 410 nm) and higher than most 2D materials (see Table S1 in the Supporting Information). Taking account of the ultrahigh $D^*$, the R-ZIS device exhibits ultrahigh sensitivity to weak near-UV signals. Actually, the photoresponse is also significant for a UV light of 370 nm, which is not included in the main text because the power cannot be measured in our lab (Figure S8 in the Supporting Information). This indicates that the R-ZIS device can also perform well in the UV spectrum. The EQE is defined as the number of electrons excited by each incident photon. Such a high EQE indicates a very high gain ($G$) of the device. When ignoring quantum efficiency ($\eta$), $G$ has the same meaning as EQE. Since $G$ is inversely proportional to the carrier transit time $\tau_T$,[53] the high $G$ means a small $\tau_T$. The EQE and $G$ are better than most 2D photoconductive materials operating under the 405 nm light.[54] Moreover, the ratio $\frac{I_{ph}}{I_{dark}}$ and LDR are calculated to be 65142 and 96.3 dB at $\lambda$ = 405 nm and $P$ = 1.64 mW cm$^{-2}$, respectively. The value of LDR is close to that of the currently used InGaAs photodetector (66 dB),[55] demonstrating that 2D R-ZIS semiconductor is a promising candidate material in low-energy-consumption optoelectronics. In addition to the present device, more devices have been tested and the data are shown in Figure S9 and S10 in the Supporting Information.

The relationship between photocurrent and excitation light intensity is important for evaluating the quality of photodetectors. Figure 5e shows the semi-logarithmic $I_{ds}$–$V_{ds}$ curves taken in the dark and under the 405 nm laser tuned to different powers. Figure 5f,g show the responsivity, specific detectivity and photocurrent as a function of laser power density calculated from the measurements at $V_{ds}$ = 5 V, respectively. The responsivity is up to 230 A W$^{-1}$ and $D^*$ is up to 1.8 × 10$^{14}$ Jones at $V_{ds}$ = 5 V and $P$ = 0.0164 mW cm$^{-2}$. The photocurrent shows an exponential dependence on the laser power, i.e., $I_{ph} \propto P^{\alpha}$, where $\alpha$ is the power exponent. The power law fit gives $\alpha$ = 0.83, which is roughly in line with the photoconductive effect.[56] Meanwhile, the sublinear relation suggests existence of trapping states that modulate the photocurrent-generation mechanism.[53] For R-ZIS, the trapping states may come from the Zn and S vacancies demonstrated by the STEM and EDX characterizations. Moreover, the device has been tested under harsh conditions to study its stability and durability. That is, the laser illumination is repeatedly turned on and off at a high frequency of 1000 Hz. Figure 5h shows 120 response cycles at $V_{ds}$ = 5 V and 3.1 mW. The ON and OFF states are well defined, stable and durable. To accommodate the fast response speed, the oscilloscope plus amplifier method has been applied to measure the response time of the device. The rising time $\tau_{rise}$ and the recovery time $\tau_{decay}$ are defined as the time between 10% and 90% of the maximum photocurrent at the rising and decaying edges, respectively. As shown in Figure 5i, the response time is determined to be $\tau_{rise}$ = 222 μs and $\tau_{decay}$ = 158 μs, indicating an ultrafast photoresponse which is superior to most other photodetectors (Table S1 in the Supporting Information). Such a fast photoresponse can be attributed to the photoconduction mechanism that induces the direct separation and recombination of electrons and holes. It is also related to the short carrier transit time $\tau_T$ as described above. This observation manifests the potential of R-ZIS in high-frequency and fast-response application scenarios.

Furthermore, an R-ZIS phototransistor with a back-gate structure has been also fabricated. Figure 6a shows a schematic diagram of the R-ZIS phototransistor. The transfer curve ($I_{ds}$–$V_{gs}$) taken without illumination indicates that R-ZIS is an n-type semiconductor with an ON/OFF current ratio of $\sim 10^5$ (Figure 6b). The electron mobility can be calculated according to the equation $\mu_e = \frac{dI_{ds}}{dV_{gs}} \times \frac{L}{W V_{ds} C_i}$, where $L$ is the channel length, $W$ is the width and $C_i$ is the capacitance per unit area (11.6 nF cm$^{-2}$ for the 300 nm SiO$_2$ layer). The field-effect electron mobility of the channel material at room temperature is calculated as 0.34 cm$^2$ V$^{-1}$ s$^{-1}$.

Figure 6b shows the transfer curves taken at $V_{ds}$ = 5 V in the dark and under the 405 nm laser tuned to different laser powers from 7.7 μW to 5.3 mW. For the ON and OFF states of the channel, the increase of gate voltage leads to the increase of photocurrent, suggesting that photocurrent is dominant over thermionic or tunneling current in the entire operating process of the R-ZIS phototransistor.[57] The dependence of responsivity, $D^*$, $I_{ph}/I_{dark}$ ratio and EQE on the gate voltage is shown in Figure 6c–f, respectively. The operation mechanism can be explained by the schematic energy band diagrams shown in Figure 6g–i. In the dark state, the device without gate voltage is in equilibrium (Figure 6g), and the carrier transport in the channel is dominated by two small contact barriers. When the channel is tuned to the OFF state under illumination ($V_{gs} < V_{th}$, $V_{th}$ is the threshold voltage), the photocurrent increases linearly with $V_{gs}$ due to the shorter carrier transit time (Figure 6h). Meanwhile, the downward shift of the Fermi level brings a higher energy barrier at the contact interface. Therefore, the photocurrent rather than the thermionic or tunneling current dominates the channel current, revealing high sensitivity to illumination. However, in the ON state under illumination ($V_{gs} > V_{th}$), the Fermi level approaches the conduction band (Figure 6i). In this way, the thermionic and tunneling current also contributes to the total current, and the photoresponse increases with the increase of $V_{gs}$ as a result of the lowering of barriers.[58] This explains why the ratio $I_{ph}/I_{dark}$ decreases as $V_{gs}$ increases, i.e., because the thermionic and tunneling current occupies a larger portion in the channel current.

In order to clarify the photocurrent-generation mechanism in the R-ZIS phototransistor, a more in-depth analysis is needed. Generally, the mechanism of 2D materials can be divided into three categories: thermal mechanism, photoconductive effect and photogating effect.[59] For our measurement, a large laser spot (5 mm in diameter) is used to cover the entire channel, so the thermal gradient is negligible and the thermal mechanism can be ignored. As shown in Figure 6b, the transfer curve shifts vertically under illumination, which can be attributed to the photoconductive effect. The photocurrent arising from the photoconductive effect leads to an increase in $I_{ds}$ through extra photo-generated carriers, which can be described as $I_{ph} = e\mu n EWD$, where $\mu$ is the carrier mobility, $n$ is the carrier density, $E$ is the electric field in the channel and $D$ is the absorption depth.[60] It has a linear dependence on the incident laser power, i.e., $I_{ph} \propto P$.[56] In addition, a horizontal shift of $V_{th}$ is observed when tuning the laser power (see the arrows in Figure 6b), which is a signature of photogating effect. If one type of photogenerated carriers is trapped in the localized states which act as a local gate, the $V_{th}$ will shift. The shift to the negative gate direction indicates that holes are trapped and induce a positive gating ($\Delta V_g > 0$). The photocurrent generated by photogating can be described as $I_{ph} = g_m \Delta V_{th}$, where $g_m$ is the transconductance and $\Delta V_{th}$ is the shift of $V_{th}$.[53] Another distinguishable feature of the photogating effect is the sublinear $I_{ph}$–$P$ relation (i.e., $I_{ph} \propto P^{\alpha}$, $\alpha < 1$).[56] Figure 7a shows the photocurrent calculated from Figure 6b as a function of laser power, with the gate voltage ranging from −80 to 70 V. The data points can be fitted to the power law, and the $\alpha$ obtained at different $V_{gs}$ is listed in Figure 7b. $\alpha$ is very close to 1 when $V_{gs} = $ −80 V or −70 V, and gradually decreases with the increase of $V_{gs}$. When $V_{gs} = 70$ V, it becomes 0.68. This indicates an evolution of the underlying mechanism from the photoconductive effect to the photogating effect modulated by the back gate. Correspondingly, the responsivity increases to about $10^4$ A W$^{-1}$ at $V_{gs} = 70$ V (Figure 7c), as a result of the enhanced photogating. Figure 7d is a schematic diagram of photogating. Under illumination, the localized states below the Fermi level can trap some photo-generated holes and cause a $\Delta V_g$. The Fermi level $E_F$ plus $\Delta V_g$ gives a larger "effective" $E_F^{'}$, thereby enhancing the photoconductivity. As $V_{gs}$ increases, more trapping states are submerged below the Fermi level, and thus the photogating effect becomes more prominent.[61]

The extraordinary performance of R-ZIS photodetector shows that it has great application potential. Here we take the application in optical neural networks as an example. As shown in Figure 8a, an optical convolution engine is designed based on the R-ZIS photodetector. The 405 nm laser with four different laser powers is used to illuminate the devices. The device is defined as being in a high resistance state (HRS) when the illumination is OFF and in a low resistance state (LRS) when the illumination is ON. The significant difference between LRS and HRS, as well as their integrated metal-semiconductor-metal structure, is the basis for the realization of the optical convolution engine. Three convolution kernels are used to demonstrate the application of R-ZIS photodetectors in ONN. As shown in Figure 8b, there are 3 × 3 pixels in each convolution kernel. The current values of the R-ZIS device in LRS and HRS are used to represent the output "1" and "0" in the binary convolutional neural network. Only when the photodetector is under illumination (weight = 1) and the input voltage is nonzero (input = 1), the output line is equal to 1 (output = 1). In all other cases, the output line is equal to zero (output = 0). The "−1" in the convolution kernels indicates that the photocurrent direction is reversed by changing the polarity of the applied voltage. This unique feature allows the implementation of both positive and negative synapses in the neural network nodes.

The calculation process of the convolutional neural network is shown in Figure 8c. The input feature map is an image of a dog selected from the CIFAR-10 dataset.[62] The 30 × 30 input feature map is transformed into the voltage amplitude vectors and then convolved with the programmed convolution kernels. The total current obtained from the convolution is recorded as one pixel in the output image. Figure 8d shows the output map images. The theoretical output maps (marked as I) are obtained by performing convolution operations on the input image and each of the three convolution kernels. In addition, the real *I–V* characteristics of the R-ZIS photodetector taken at different powers are combined with the kernels. For a low laser power (7.7 μW), the output maps (II) fit the theoretical outputs well. The correlation coefficient (> 82%) is acceptable (see Table S2 in the Supporting Information). When the laser power exceeds 0.41 mW, the output maps (III) can highly reproduce the theoretical outputs, and the correlation coefficient becomes higher (> 94%). For the laser power of 4.30 mW, the correlation coefficient can reach 98%. These results indicate that the construction of the optical convolution engine based on the R-ZIS photodetectors is successful and confirm the effectiveness of the application in image computation. A ready-to-use application scenario is deep neural networks. In many cases, deep neural network models are intellectual property rights that need to be protected.[63] Digital watermarking technology is used for protection, which requires a computing system that does not store weight data but securely receives them during inferences.[64] The R-ZIS optical convolution engine may be an ideal device for this purpose.

## CONCLUSIONS

In summary, we have grown high-quality rhombohedral ZIS single crystals and performed detailed characterizations to study the crystal structure, chemical composition and electronic structure. The photodetectors based on the exfoliated R-ZIS nanosheets exhibit a broadband photoresponse from near UV to visible light. The superior performance represented by a series of parameters, i.e., extremely low dark current (7 pA at 5 V bias), specific detectivity ($1.8 \times 10^{14}$ Jones), responsivity (230 A W$^{-1}$), response time ($\tau_{\text{rise}} = 222$ µs, $\tau_{\text{decay}} = 158$ µs) and EQE ($6.12 \times 10^4$%) for the 405 nm laser, surpasses most 2D counterparts. Moreover, the back gate can effectively modulate the photocurrent-generation mechanism from the photoconductive effect to the dominant photogating, and increases the responsivity by two orders of magnitude ($10^4$ A W$^{-1}$). The combination of ultrahigh sensitivity, ultrafast response and high gate tunability makes the R-ZIS phototransistor an ideal device for low-energy-consumption and high-frequency optoelectronic applications, which is further highlighted by its excellent performance in ONN.

**ASSOCIATED CONTENT**

**Supporting Information**

The Supporting Information is available free of charge at https://xxx.

Schematic diagram of CVT growth of R-ZIS single crystals and an optical image; powder XRD pattern; optical image showing more R-ZIS flakes; SAED pattern and HRTEM image taken on the cross section of an R-ZIS thin flake prepared by focused ion beam; STEM-EDX spectrum; optical image and *I–V* curves of R-ZIS photodetector based on Cr/Au contact; time-resolved photocurrent of R-ZIS photodetector taken under lasers of different wavelength; *I–V* curve taken under illumination of 370 nm; test results of more devices (S2 and S3); diagram of optoelectronic measurement setup; comparison of photoresponse characteristics of R-ZIS with some reported photodetectors based on 2D materials; correlation coefficients between the theoretical results and the outputs combining R-ZIS photodetector *I–V* characteristics; reliability of using TEM to identify atoms in thin flakes; noise current and dark current

## Author Contributions

The manuscript was prepared through the contributions of all authors.

## Notes

The authors declare no competing financial interest.

## ACKNOWLEDGMENTS

We appreciate the discussion with Professor Liang Zhao. This work was supported by the National Key R&D Program of China (Grant No. 2021YFA1600201), the National Natural Science Foundation of China (Grant No. 11874363, 11974356 and U1932216) and Anhui Province Laboratory of High Magnetic Field (Grant No. AHHM-FX-2020-01).

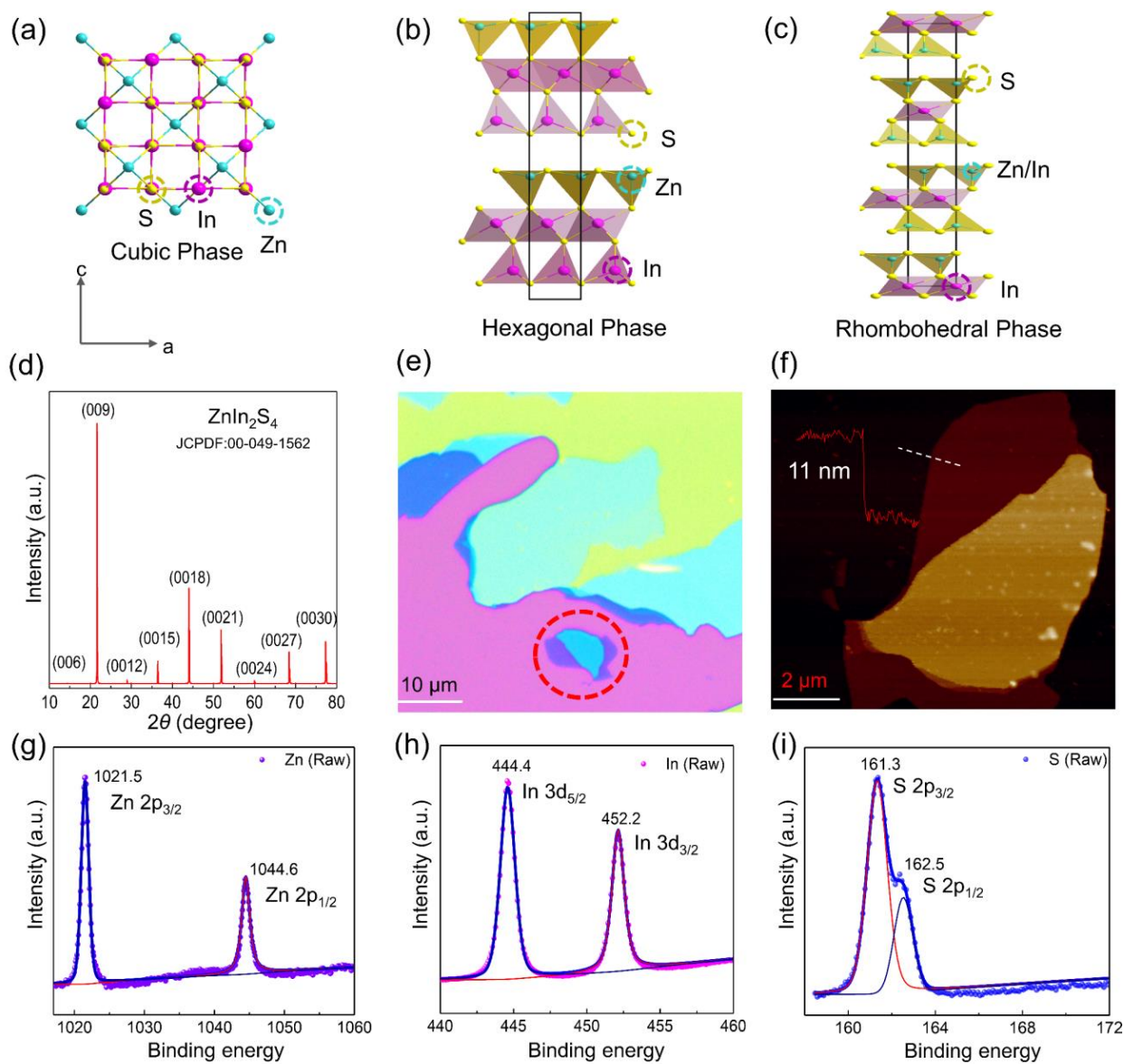

**Figure** 1. (a)-(c) Schematic diagrams of different crystal structures of ZIS, namely (a) cubic phase, (b) hexagonal phase and (c) rhombohedral phase. The green, purple and yellow balls represent Zn, In and S atoms, respectively. Note that the cubic phase is not layered, while the hexagonal and rhombohedral phases are layered. (d) Single crystal XRD pattern of R-ZIS, showing out-of-plane orientation. (e) Typical optical image of exfoliated R-ZIS flakes on SiO₂/Si substrate. (f) AFM image of an 11 nm thick R-ZIS nanosheet, corresponding to three layers. The height profile is extracted along the white dashed line. (g)-(i) Zn 2*p*, In 3*d* and S 2*p* branches of the XPS spectrum of R-ZIS nanosheets. The solid curves represent the background, the total and individual fitting results of peak analysis.

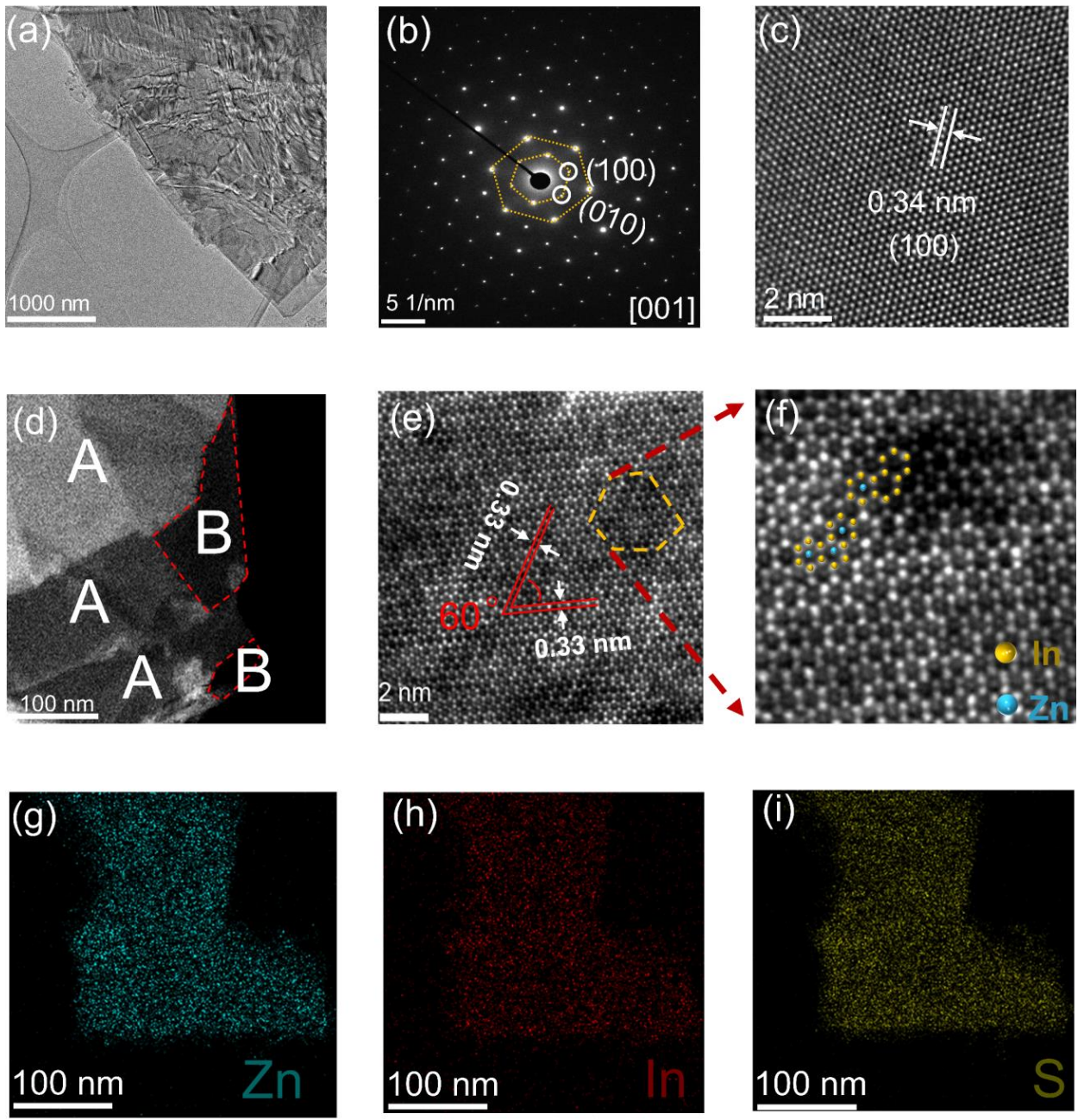

**Figure** 2. (a) Low-magnification HAADF-STEM image of a typical R-ZIS nanosheet. (b) SAED pattern of the R-ZIS flake corresponding to (a). The white circles highlight the (100) and (010) planes. The zone axis is [001]. (c) High-resolution TEM (HRTEM) image of (a). (d) Low magnification HAADF-STEM image of another R-ZIS nanosheet. A and B denote few-layer zone and single-layer zone, respectively. (e) The corresponding atomic-resolution HAADF-STEM image of the marked region in (d). (f) An enlarged view of the area circled in (e) clearly showing a certain number of Zn vacancies. The cyan and yellow balls represent Zn and In atoms, respectively. (g)-(i) are the EDX mapping results for Zn, In and S, respectively.

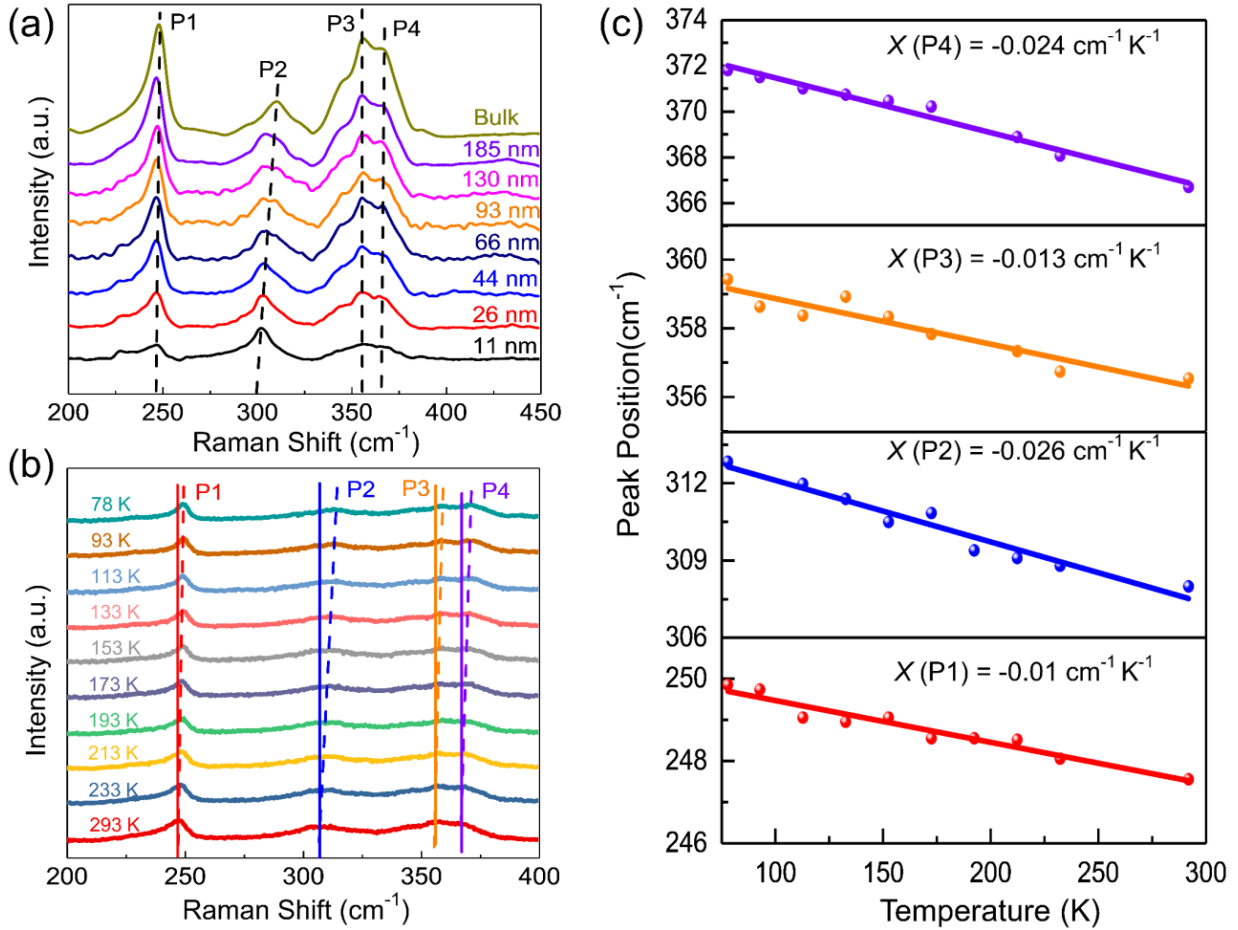

**Figure** 3. (a) Raman spectra of R-ZIS of various thicknesses from bulk to three layers taken at room temperature. The dashed lines indicate the peak positions. (b) Raman spectra of the 44 nm thick R-ZIS flake measured at different temperatures from 78 K to 293 K. The dashed lines indicate the peak positions, and the solid lines are the reference (293 K). (c) Peak positions of P1, P2, P3 and P4 as a function of temperature. The solid lines are the linear fits.

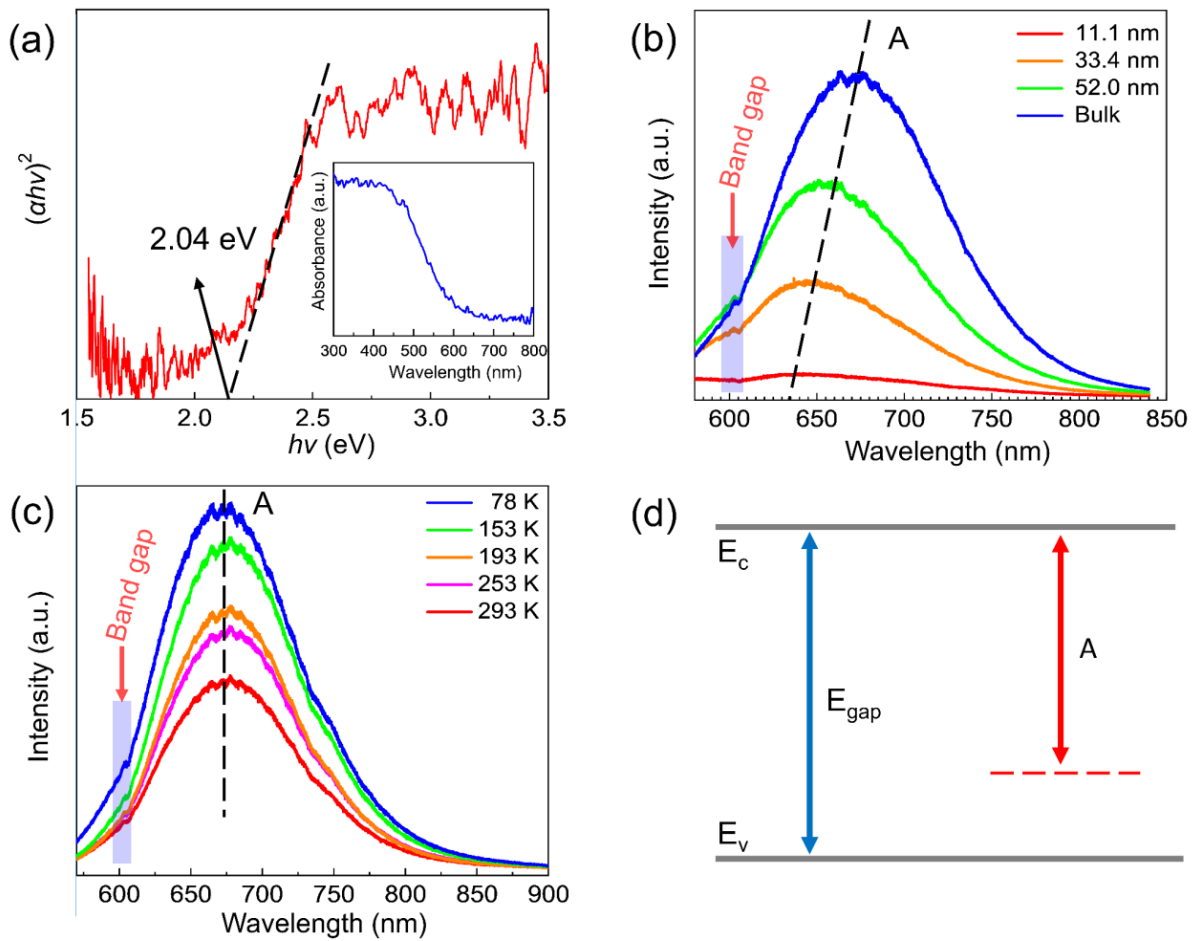

**Figure** 4. (a) Absorption spectrum of R-ZIS single crystal plotted as $(\alpha h\nu)^2$ vs. photon energy $h\nu$. Inset: Raw data taken from UV to visible light. (b) PL spectra of R-ZIS flakes of different thicknesses ranging from bulk to three layers. The light blue vertical bar indicates the position of the bandgap. (c) PL spectra of the 44 nm thick R-ZIS nanosheet taken at various temperatures from 78 K to 293 K. (d) Schematic diagram of energy levels. The dashed line represents the doping level due to defects.

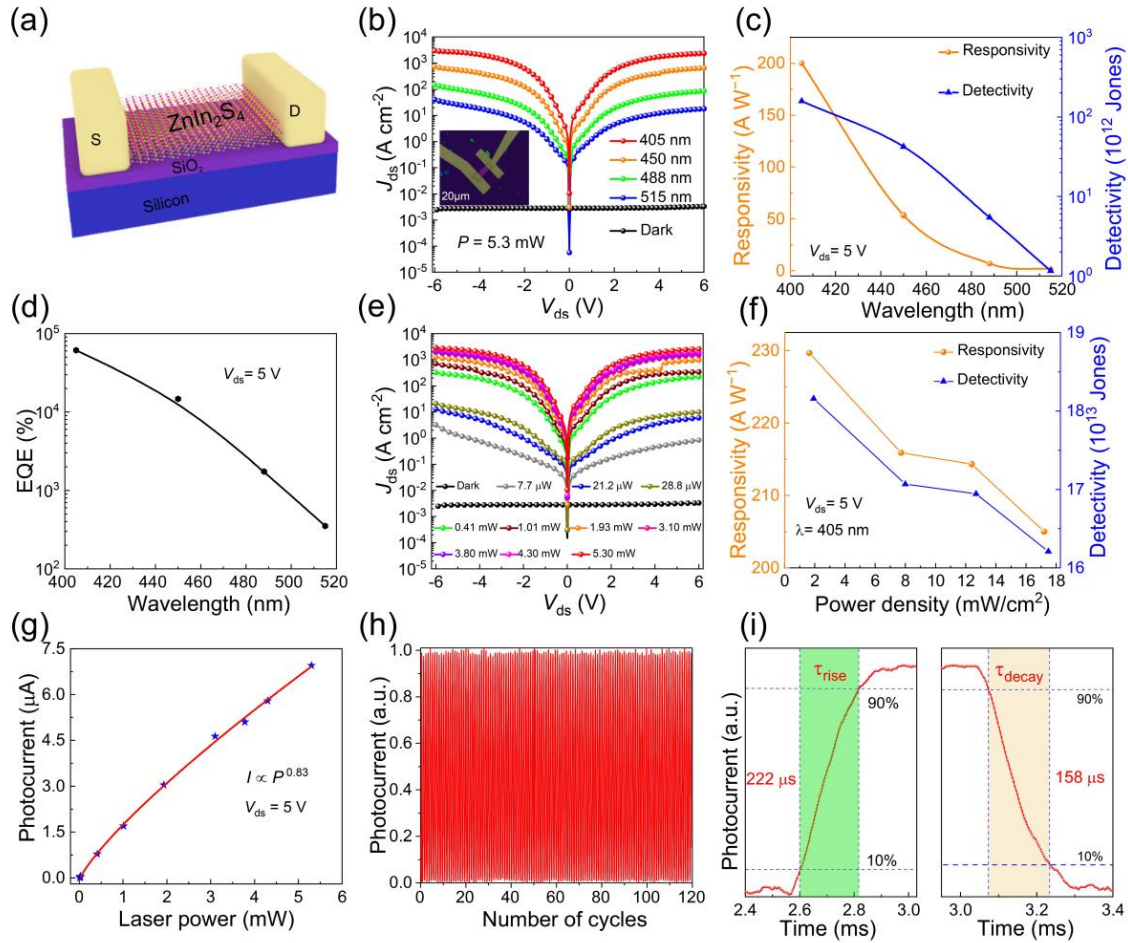

**Figure** 5. (a) Schematic diagram of R-ZIS photodetector. (b) Current–voltage curves plotted on a semi-logarithmic scale taken in the dark and under laser illumination of different wavelengths. Inset: Optical image of the photodetector based on an 18 nm thick R-ZIS flake. The scale bar is 20 μm. (c) Responsivity (left), specific detectivity (right) and (d) EQE of the R-ZIS photodetector plotted as a function of laser wavelength at $V_{ds}$ = 5 V. (e) Current–voltage curves taken in the dark and under a 405 nm laser tuned to different powers. (f) Dependence of responsivity (left) and specific detectivity (right) on laser power density at $V_{ds}$ = 5 V. (g) Photocurrent at $V_{ds}$ = 5 V as a function of laser power. The curve represents a power law fit. (h) 120 response cycles of the R-ZIS photodetector setting the laser power to 3.1 mW. (i) An enlarged view of a single response cycle, including the rising and decaying processes.

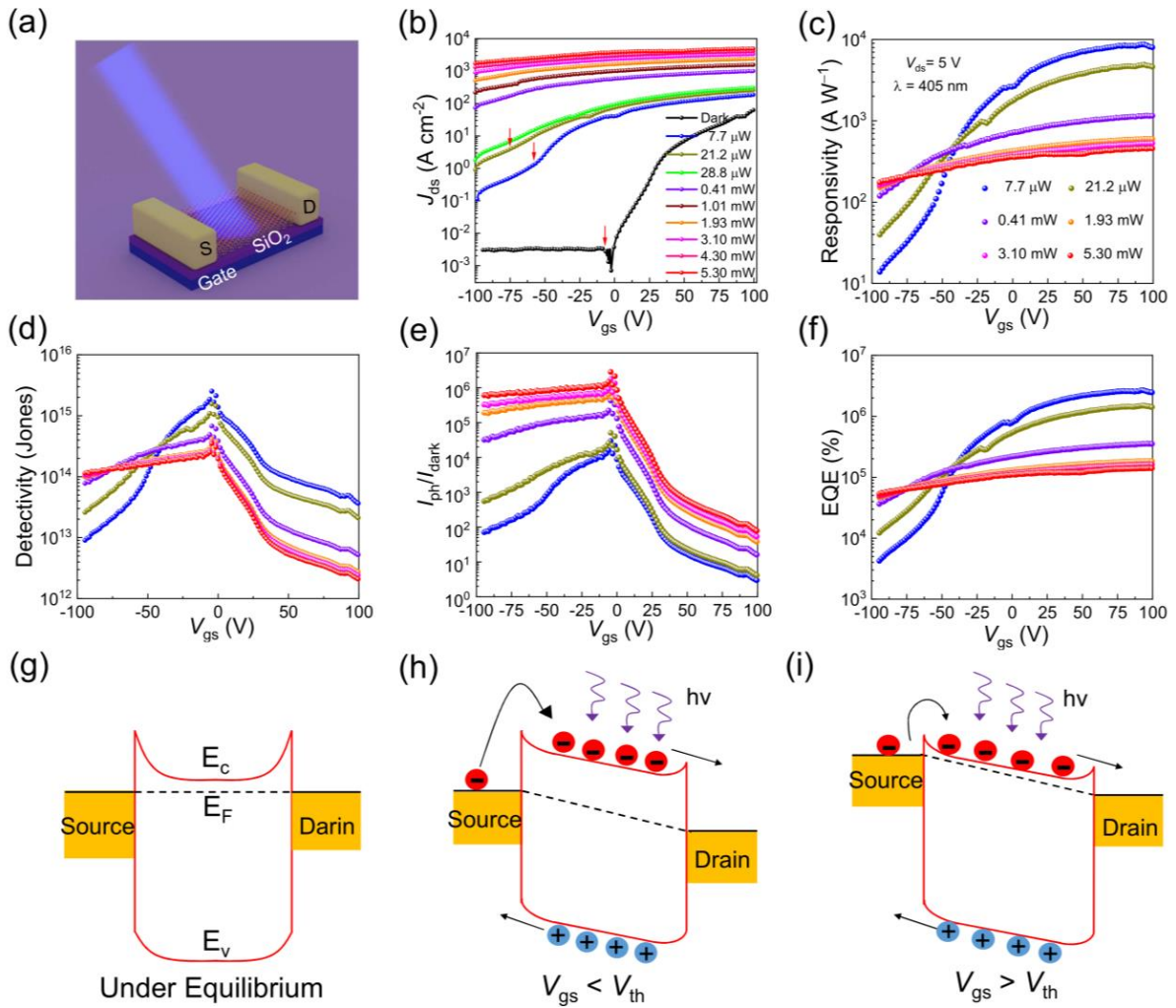

**Figure** 6. (a) Schematic diagram of R-ZIS phototransistor. (b) Transfer curves taken at $V_{ds}$ = 5 V in the dark and under the 405 nm laser tuned to different laser powers. The arrows indicate the threshold voltages. (c) Responsivity, (d) specific detectivity, (e) $I_{ph}/I_{dark}$ and (f) EQE as a function of $V_{gs}$ calculated for $V_{ds}$ = 5 V and different laser powers. (b)-(f) share the same color configuration. (g)-(i) Energy band diagram of R-ZIS phototransistor in different states. $E_F$, $E_c$, and $E_v$ represent the Fermi level, conduction band minimum and valence band maximum, respectively.

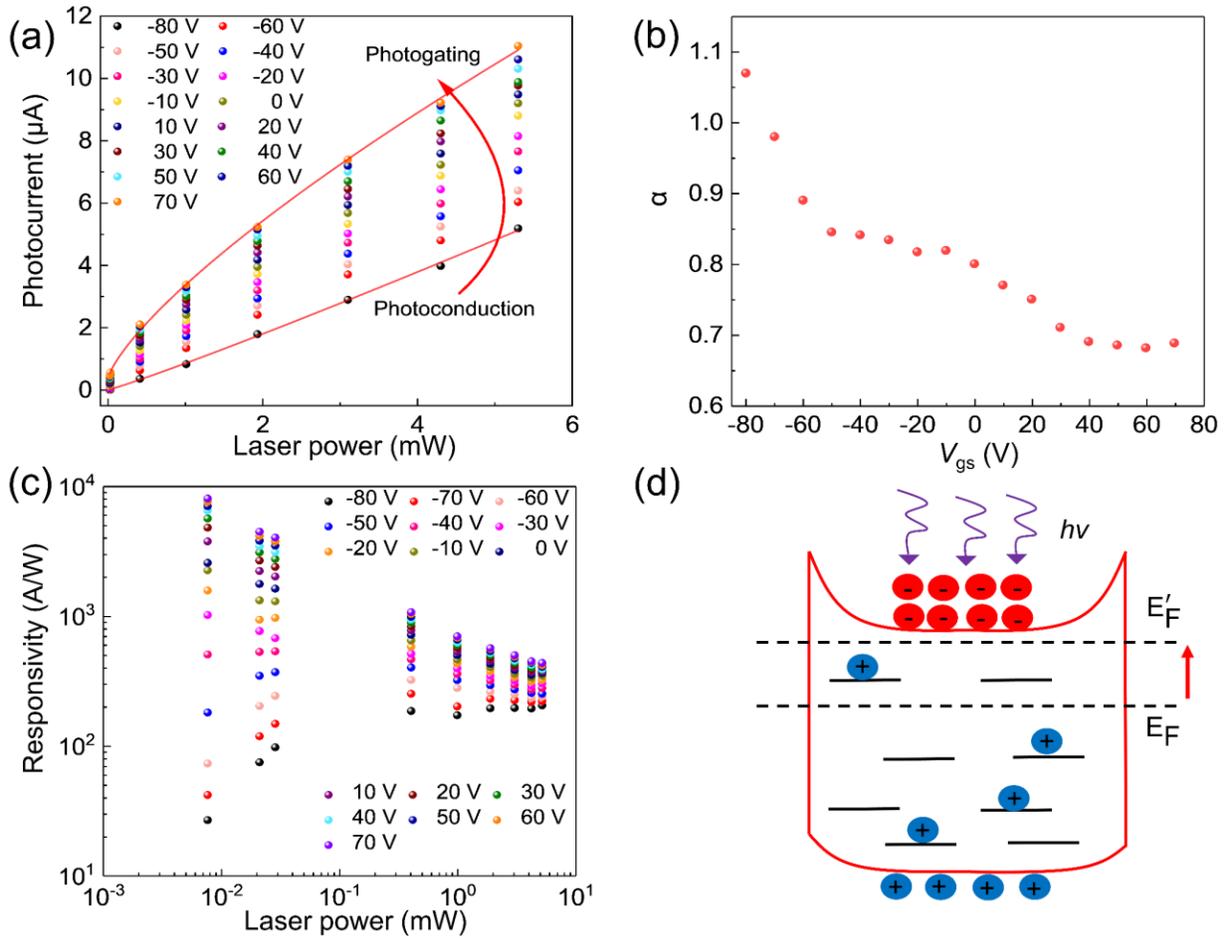

**Figure** 7. (a) Photocurrent (calculated from Figure 6b) as a function of incident laser power with different gate voltages. (b) $\alpha$ extracted from (a) for different gate voltages. (c) Responsivity (calculated from Figure 6b) as a function of incident laser power with different gate voltages. (d) Schematic diagram of the photogating effect.

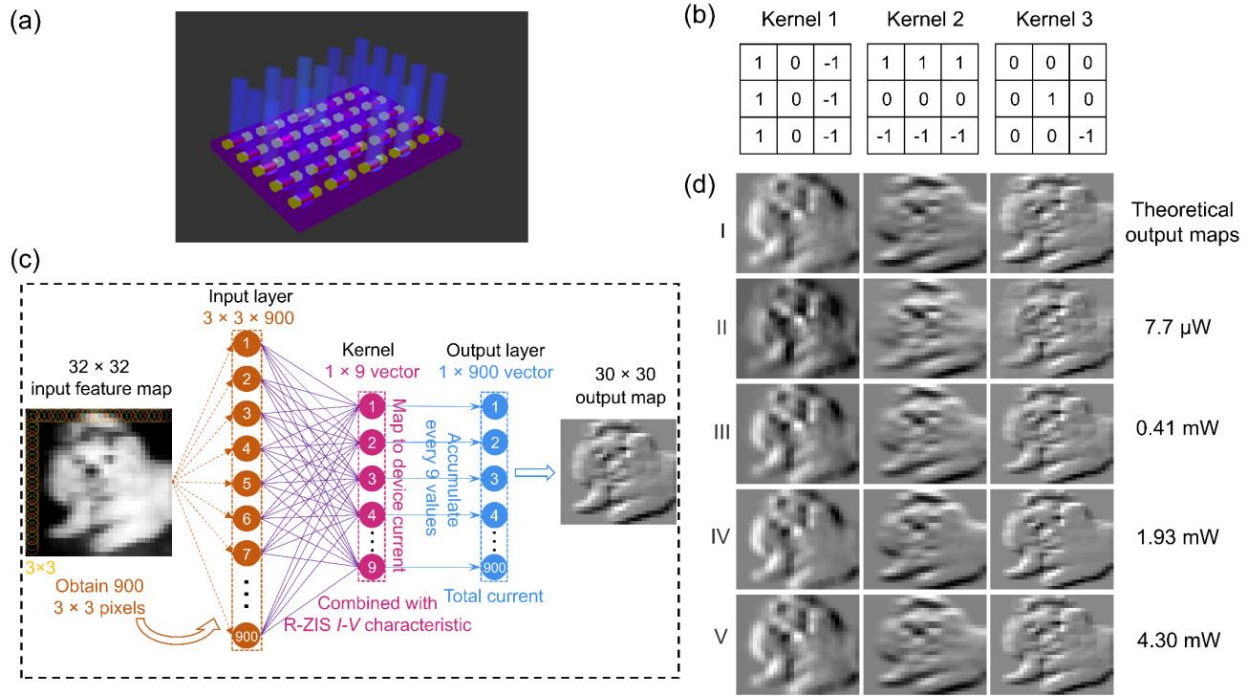

**Figure** 8. (a) Schematic illustration of optical convolution engine based on R-ZIS devices. (b) Three different convolution kernels with 3 × 3 pixels. (c) Scheme of convolutional neural network. The R-ZIS photodetector model provides an idea for the design of the optical convolution engine, in which the input map is converted into different voltage amplitudes and the significant current difference between the laser illumination and the dark state under the same input voltage. The input feature image is selected from the CIFAR-10 dataset. (d) The theoretical output maps after convolution operation using the input maps and three kernels are marked as I. The output maps using three kernels combined with the R-ZIS photodetector characteristic are marked as II (the laser power is 7.7 μW), III (0.41 mW), IV (1.93 mW) and V (4.30 mW).

TOC

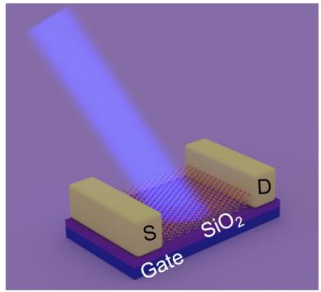

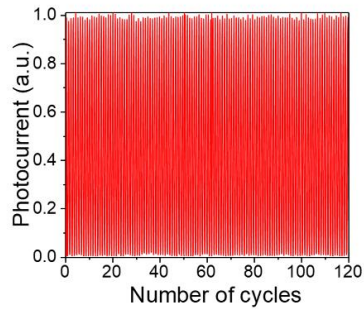

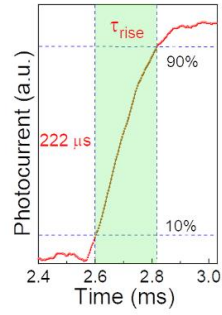

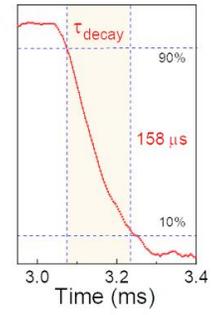

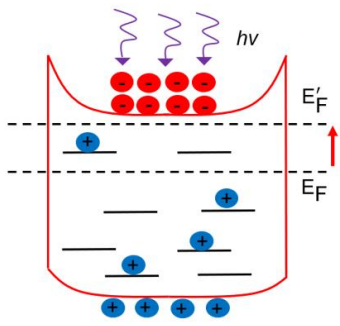

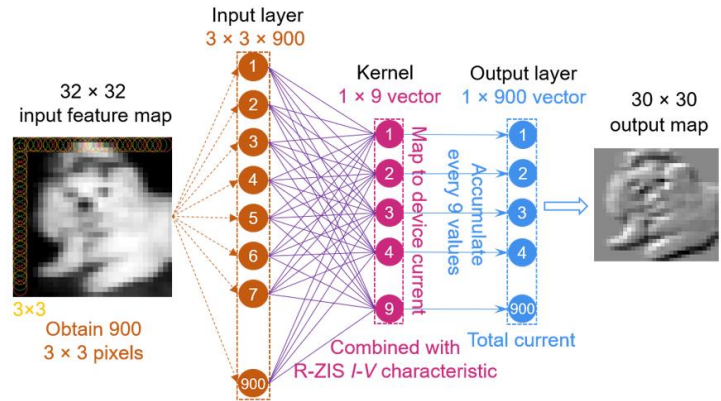

# Supporting Information

**Ultrasensitive, Ultrafast and Gate-Tunable Two-Dimensional Photodetectors in Ternary Rhombohedral ZnIn₂S₄ for Optical Neural Networks**

Weili Zhen[a,b,§], Xi Zhou[d,e,§], Shirui Weng[a], Wenka Zhu[a,*], and Changjin Zhang[a,c,†]

*[a]High Magnetic Field Laboratory, Chinese Academy of Sciences, Hefei 230031, China*

*[b]University of Science and Technology of China, Hefei 230026, China*

*[c]Institutes of Physical Science and Information Technology, Anhui University, Hefei 230601, China*

*[d]The Interdisciplinary Research Center, Shanghai Advanced Research Institute, Chinese Academy of Sciences, Shanghai 201210, China*

*[e]School of Microelectronics, University of Chinese Academy of Sciences, Beijing 100049, China*

*Email: wkzhu@hmfl.ac.cn

†Email: zhangcj@hmfl.ac.cn

§W.Z. and X.Z. contributed equally to this paper

# 1. Figures and tables

## 1.1. Schematic diagram of CVT growth of R-ZIS single crystals and an optical image

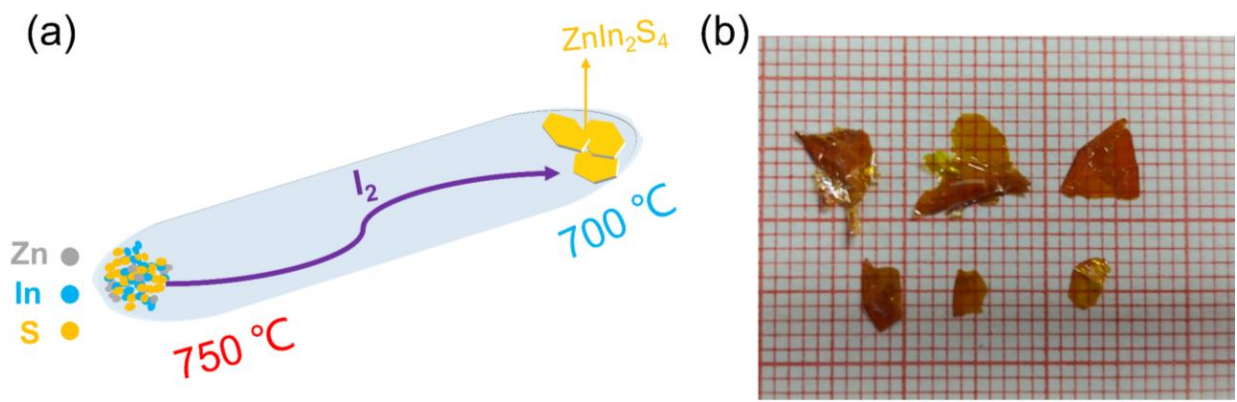

**Figure S1.** (a) Schematic diagram of chemical vapor transport growth of R-ZIS sing crystals. (b) An optical image of the as-grown R-ZIS single crystals. The size of each small grid is 1×1 mm².

## 1.2. Powder XRD pattern

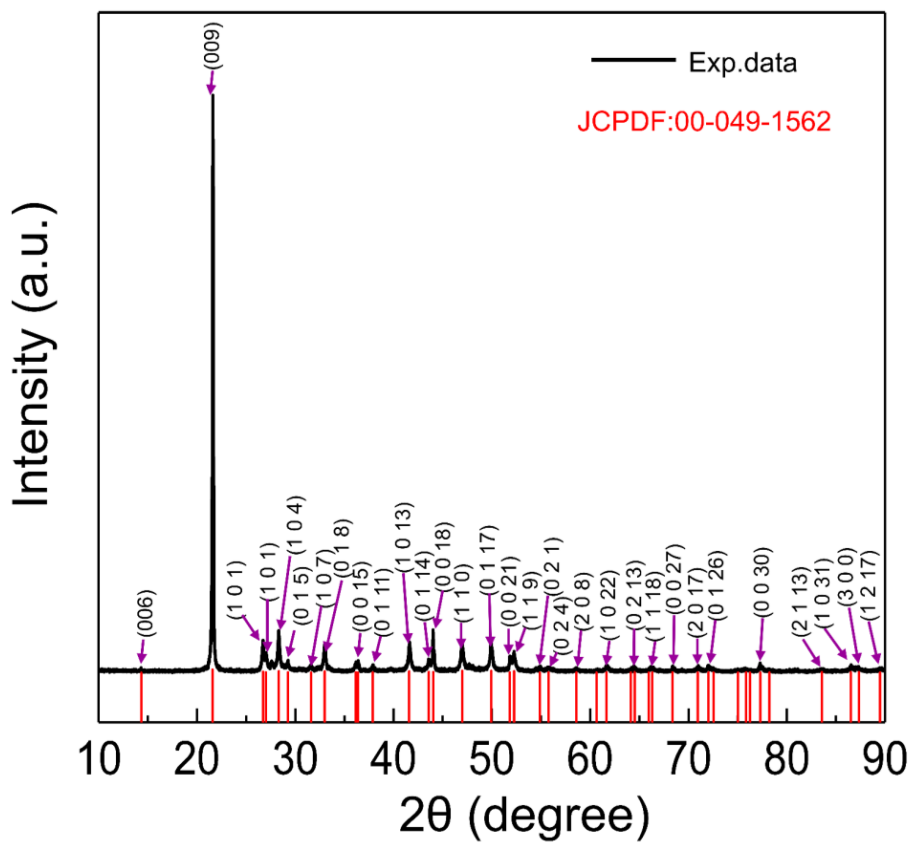

**Figure S2.** Powder XRD pattern of R-ZIS. The red ticks indicate the positions of the diffractions

of PDF No. 00-049-1562.

## 1.3. Optical image showing more R-ZIS flakes

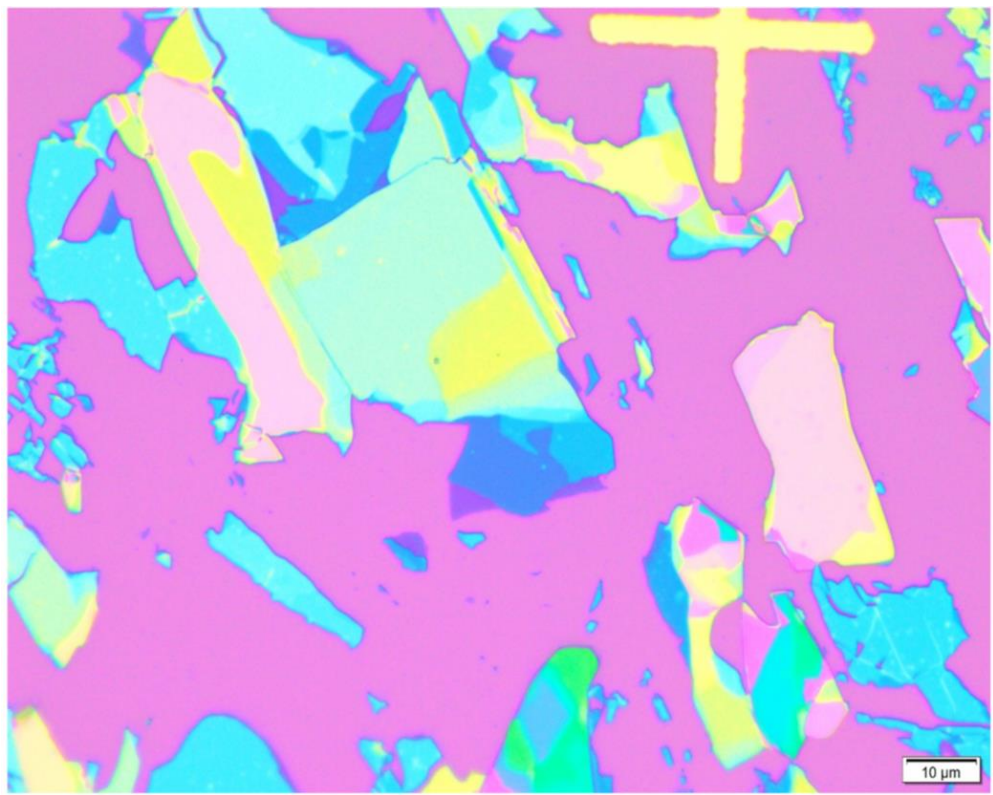

**Figure S3.** Optical image of R-ZIS flakes on SiO$_2$/Si substrate. Different optical contrasts correspond to different thicknesses and layers.

**1.4. SAED pattern and HRTEM image taken on the cross section of an R-ZIS thin flake prepared by focused ion beam**

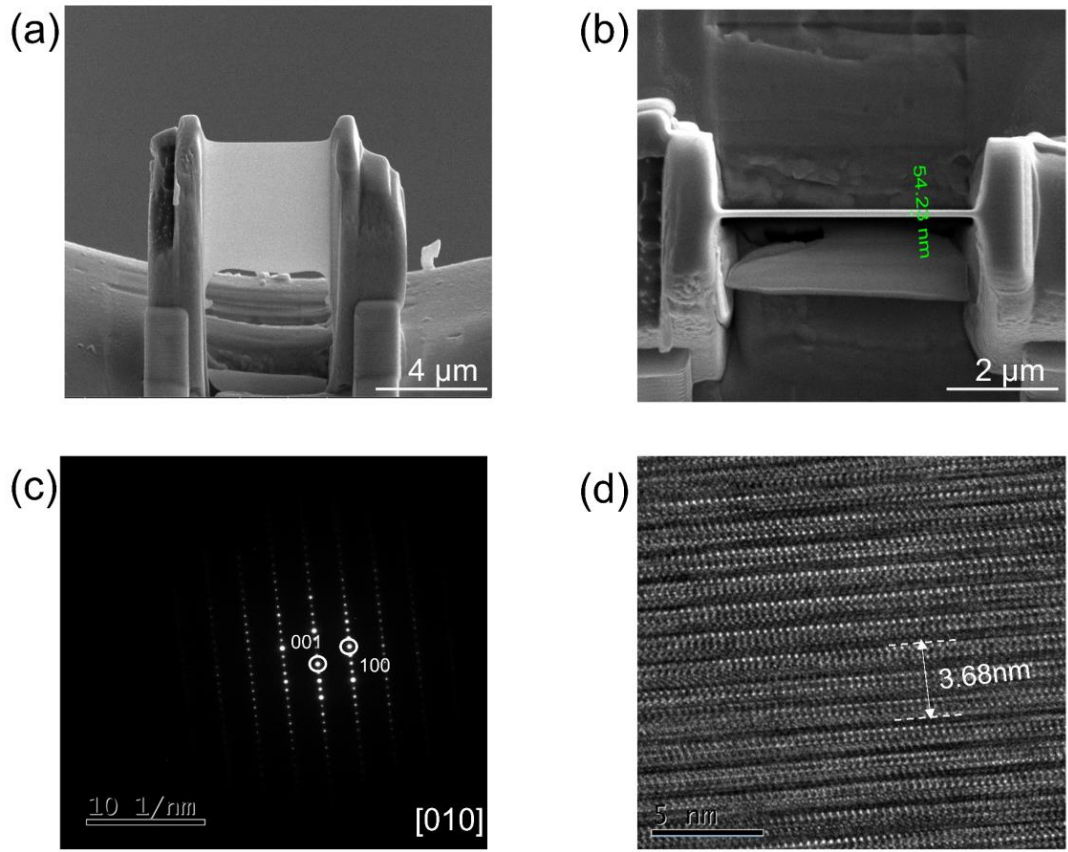

**Figure S4.** (a) R-ZIS thin flake prepared by focused ion beam. (b) Cross-sectional view of the R-ZIS sheet. The thickness is about 54 nm. (c) SAED pattern taken on the sample shown in (a) and (b). The diffraction spots correspond to the [001] and [100] crystallographic planes. The zone axis is [010]. (d) HRTEM image. The interplanar spacing is determined to be about 3.68 nm, which is close to the lattice parameter $c$ = 3.7 nm.

## 1.5. STEM-EDX spectrum

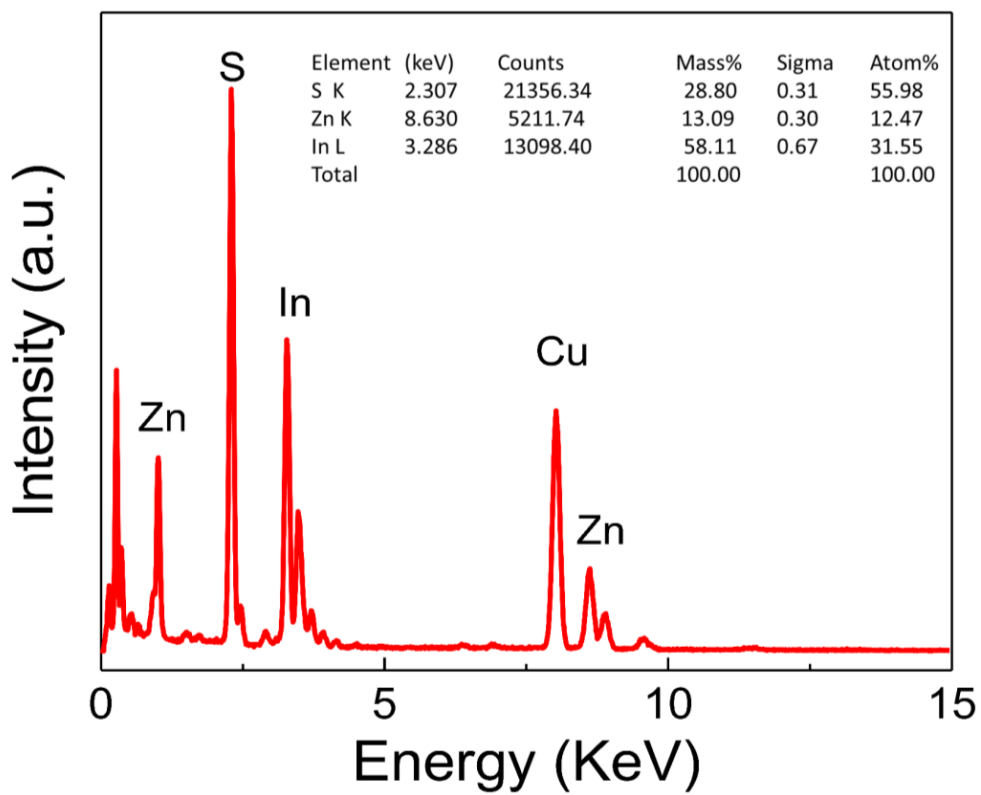

| Element | (keV) | Counts | Mass% | Sigma | Atom% |
|---------|-------|--------|-------|-------|-------|
| S K | 2.307 | 21356.34 | 28.80 | 0.31 | 55.98 |
| Zn K | 8.630 | 5211.74 | 13.09 | 0.30 | 12.47 |
| In L | 3.286 | 13098.40 | 58.11 | 0.67 | 31.55 |
| Total | | | 100.00 | | 100.00 |

**Figure S5.** STEM-EDS spectrum of R-ZIS flake. The atomic ratio is close to 1:2:4.

## 1.6. Optical image and *I–V* curves of R-ZIS photodetector based on Cr/Au contact

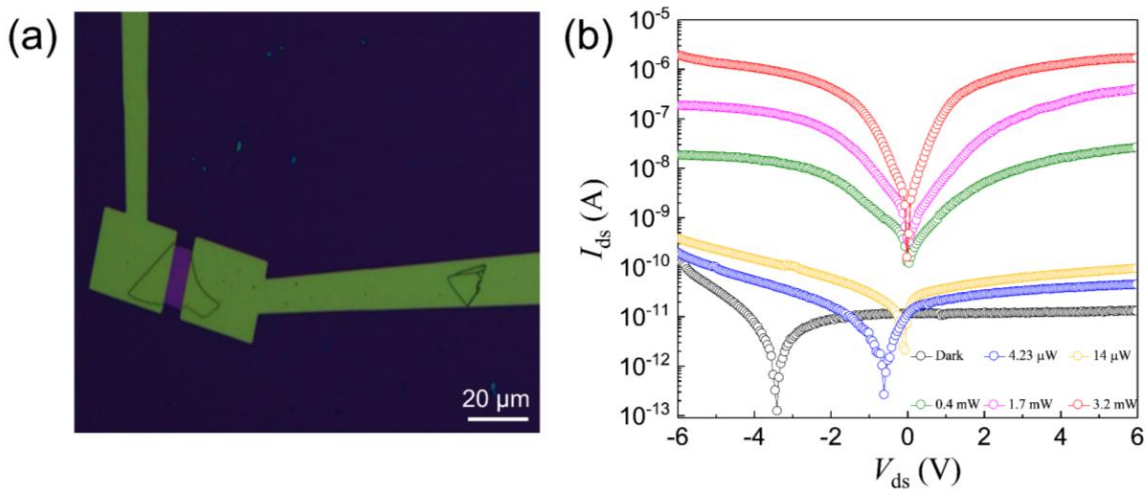

**Figure S6.** (a) Optical image of R-ZIS photodetector based on Cr/Au contact. The scale bar is 20 μm. (b) $I_{ds}$–$V_{ds}$ characteristics taken in the dark and under a 405 nm laser tuned to different laser

powers.

**1.7. Time-resolved photocurrent of R-ZIS photodetector taken under lasers of different wavelength**

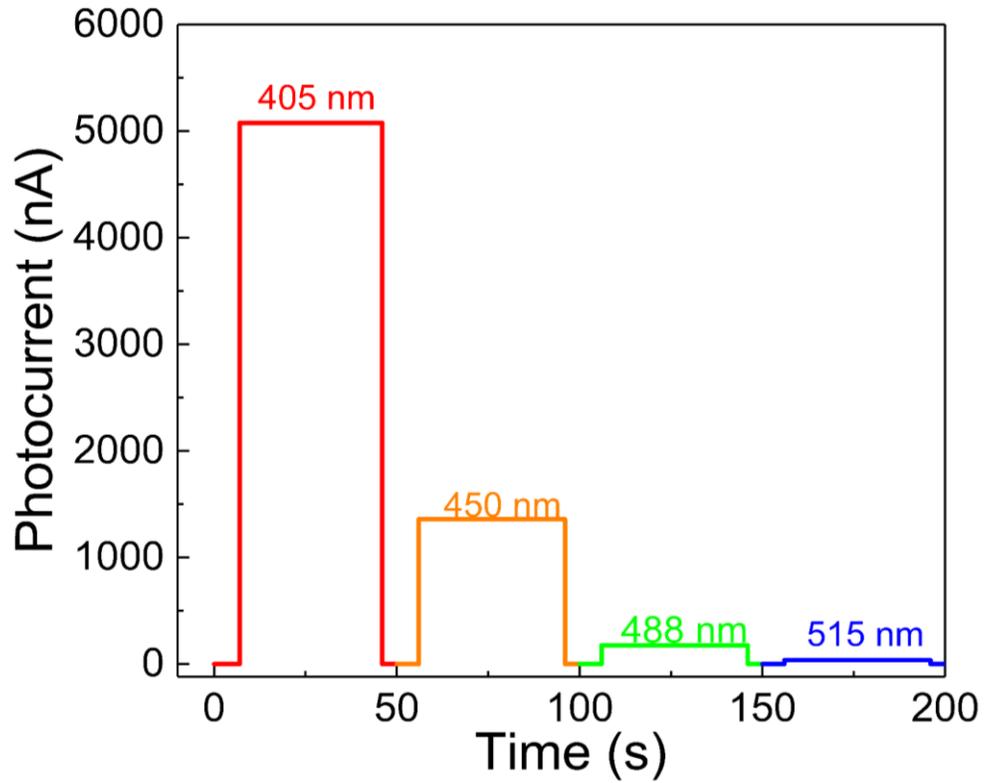

**Figure S7.** Time-resolved photocurrent of R-ZIS photodetector taken under lasers of different wavelength (405 nm, 450 nm, 488 nm and 515 nm) at $V_{ds}$ = 5 V and 5.3 mW.

**1.8. *I–V* curve taken under illumination of 370 nm**

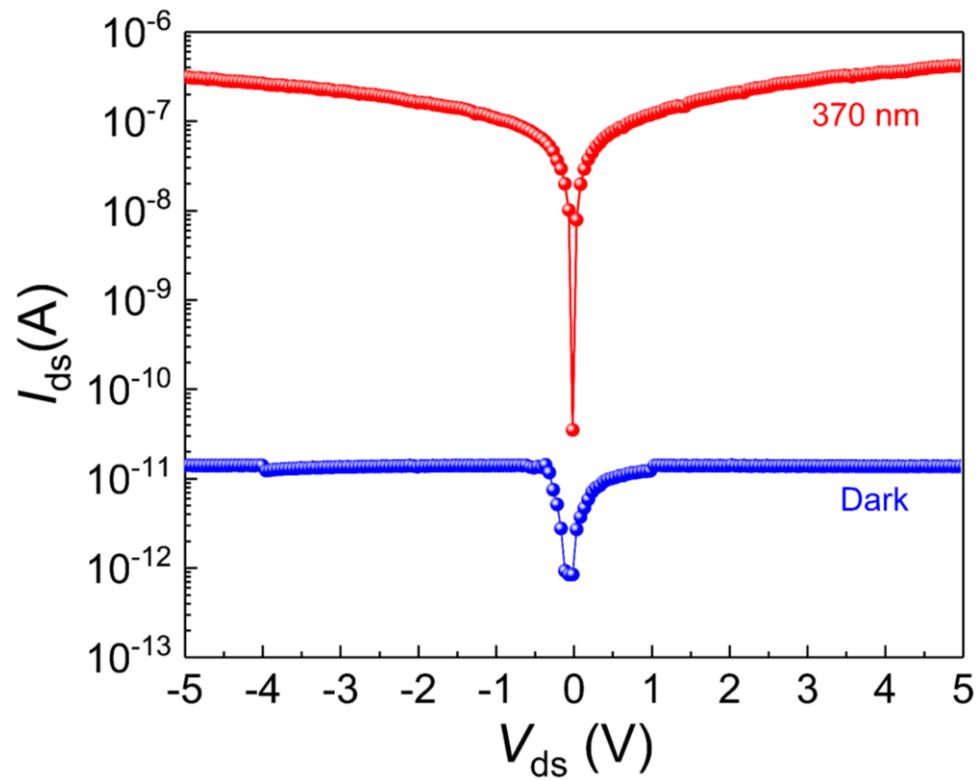

**Figure S8.** Semi-logarithmic $I_{ds}$–$V_{ds}$ characteristics taken in the dark and under illumination of 370 nm.

## 1.9. Test results of more devices (S2 and S3)

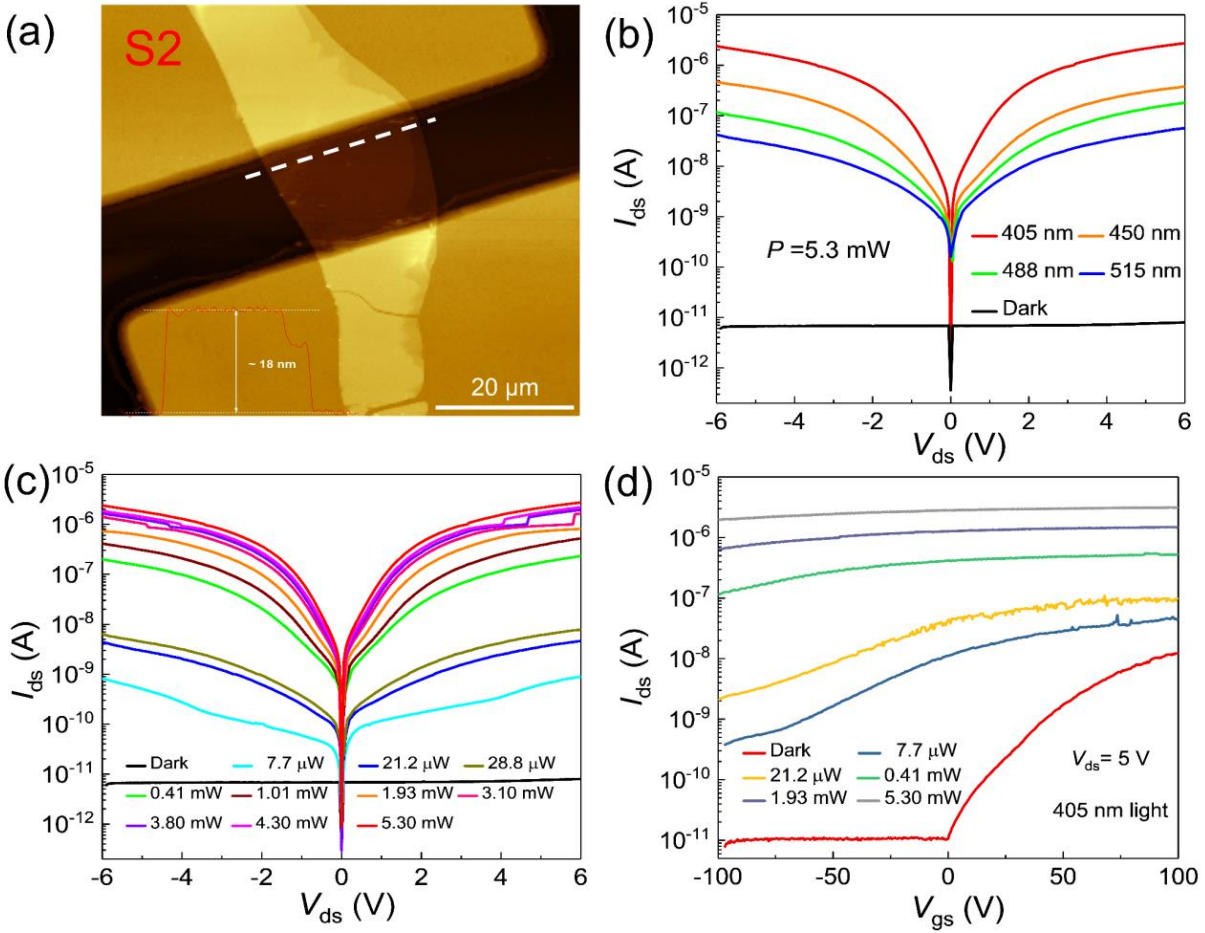

**Figure S9.** (a) AFM image of device S2. The thickness of R-ZIS flake is about 18 nm. The scale bar is 20 μm. (b) Semi-logarithmic $I_{ds}$–$V_{ds}$ curves taken in the dark and under illumination of different wavelengths. The laser power is tuned to 5.3 mW. (c) $I_{ds}$–$V_{ds}$ curves taken in the dark and under a 405 nm laser tuned to different powers. (d) Transfer curves at $V_{ds} = 5$ V under the 405 nm laser.

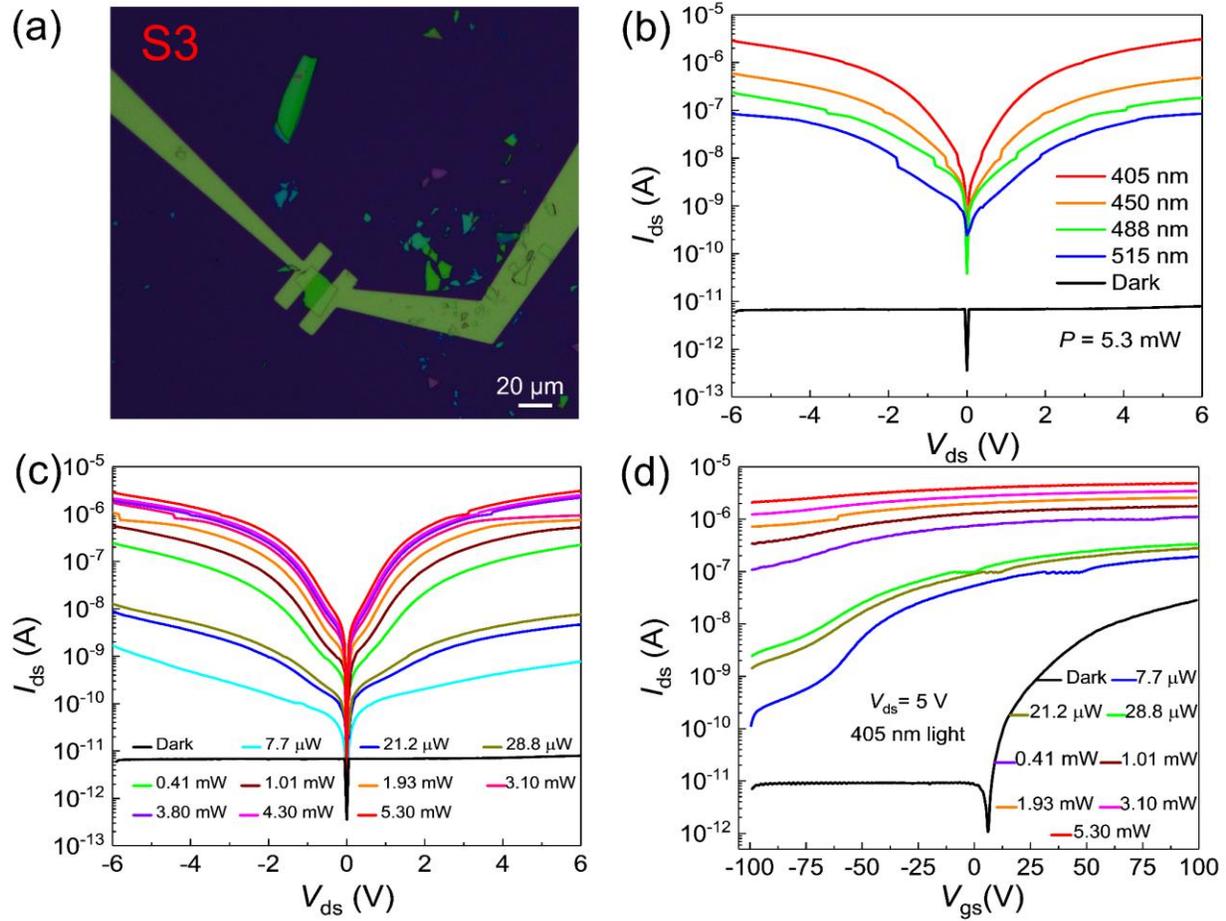

**Figure S10.** (a) Optical image of device S3. The scale bar is 20 µm. (b) Semi-logarithmic $I_{ds}$–$V_{ds}$ curves taken in the dark and under illumination of different wavelengths. The laser power is tuned to 5.3 mW. (c) $I_{ds}$–$V_{ds}$ curves taken in the dark and under a 405 nm laser tuned to different powers. (d) Transfer curves at $V_{ds}$ = 5 V under the 405 nm laser.

## 1.10. Diagram of optoelectronic measurement setup

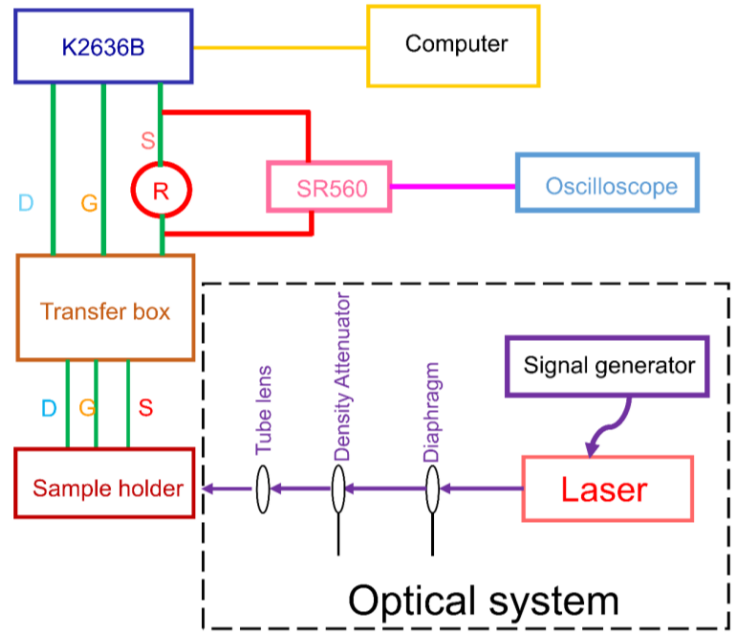

**Figure S11.** Diagram of optoelectronic measurement setup. The incident laser is produced by a fiber-coupled laser and the frequency of the laser can be changed by an external arbitrary signal generator (Tektronix AFG31021). The power of illumination laser is determined by a power and energy meter (Laser Power Meter LP10, Sanwa), and electrodes of the device are connected to a semiconductor device analyzer (Keithley 2636B) equipped with a homemade transfer box. The current signal is converted to a voltage signal using a resistance box, and then the voltage signal is passed to the preamp (Stanford Research System, SR560). Finally, the output voltage (from SR560) is displayed on an oscilloscope (Tektronix MDO3014).

## 1.11. Comparison of photoresponse characteristics of R-ZIS with some reported photodetectors based on 2D materials

**Table S1.** Comparison of photoresponse characteristics of ZIS with some reported photodetectors based on 2D materials.

| Material | Laser (nm) | $V_{bias}$ (V) | $V_g$ (V) | $I_{dark}$ (nA) | $R$ (A W$^{-1}$) | $D^*$ (Jones) | $\tau_{rise}$ (ms) | $\tau_{decay}$ (ms) |
|---|---|---|---|---|---|---|---|---|
| InSe[1] | 450 | 10 | 70 | 2000 | 157 | $<10^{12}$ | 50 | $4\times10^3$ |
| ReSe$_2$[2] | 633 | 0.5 | 0 | 2.5 | 95 | - | 68 | 34 |
| SnS$_2$[3] | 350 | 1 | 20 | 525 | 400 | - | 20 | 16 |
| GaSe[4] | 380 | 4 | 0 | 0.5 | 2.6 | $1.0\times10^{12}$ | $2.37\times10^4$ | $6.3\times10^3$ |
| Si$_2$Te$_3$[5] | 450 | 1 | 40 | 0.08 | 27 | $2.8\times10^{12}$ | 210 | 478 |
| NiPS$_3$[6] | 245 | 10 | 0 | $<10^{-5}$ | 0.13 | $1.2\times10^{12}$ | 3.2 | 17.2 |
| MnPS$_3$[7] | 365 | 8 | 70 | $1.4\times10^{-3}$ | 288 | $6.48\times10^{11}$ | 340 | 600 |
| Ta$_2$NiSe$_5$[8] | 808 | 0.1 | 0 | 4500 | 17.21 | - | 3000 | 3300 |
| Ga$_2$In$_4$S$_9$[9] | 360 | 5 | 0 | 0.5 | 112 | $2.25\times10^{11}$ | 40 | 50 |
| R-ZnIn$_2$S$_4$ (This work) | 405 | 5 | 0 | $7\times10^{-3}$ | 230 | $1.8\times10^{14}$ | 0.22 | 0.158 |
| R-ZnIn$_2$S$_4$ (This work) | 405 | 5 | 70 | 66 | $1.0\times10^4$ | $3.7\times10^{13}$ | 0.75 | 0.536 |

### 1.12. Correlation coefficients between the theoretical results and the outputs combining R-ZIS photodetector *I–V* characteristics

**Table S2.** Correlation coefficients between the theoretical results and the outputs combining ZIS photodetector *I–V* characteristics.

| Map number | Kernel 1 | Kernel 2 | Kernel 3 |
|---|---|---|---|
| II | 0.8192 | 0.9175 | 0.8773 |
| III | 0.9449 | 0.9655 | 0.9564 |
| IV | 0.9598 | 0.9769 | 0.9648 |
| V | 0.9677 | 0.9813 | 0.9767 |

### 2. Reliability of using TEM to identify atoms in thin flakes

Aberration-corrected scanning transmission electron microscopy (STEM) is an effective tool for characterizing thin-flake samples at the atomic scale. In previous researches, few-layer or monolayer 2D materials were indeed unstable or even damaged under high-energy electron beams with acceleration voltages over 120 kV [PRL, 109, 035503 (2012), Ultramicroscopy 146, 33 (2014)]. However, by lowering the acceleration voltage, typically below 100 kV, the damage caused by the electron beam to the materials can be reduced, allowing the identification of atoms and defects in thin materials. This has been achieved in many few-layer and monolayer 2D materials, such as $MoS_2$ [Adv. Mater. 26, 2857 (2014), Adv. Mater. 26, 2648 (2014)], $WSe_2$ [Chem. Mater. 33, 1307 (2021), Adv. Mater. 2022, 2106551, Nat. Phys. 17, 92 (2021)], $MoTe_2$ [ACS Nano 11, 11005 (2017)], InSe [ACS Nano 13, 5112 (2019)], GaSe [ACS Nano 9, 8078 (2015)] and $ReS_2$ [ACS Nano 9, 11249 (2015)]. These results demonstrate the reliability of this identification.

In addition, for the monolayer and few-layer hexagonal phase $ZnIn_2S_4$, an isomer of rhombohedral $ZnIn_2S_4$ in the present work, the identification of atoms and Zn vacancies by STEM has been successfully performed [JACS 139, 7586 (2017), ACS Nano 15, 15238 (2021),

ChemSusChem 14, 852 (2021)]. Based on these experiences, we chose a relatively low acceleration voltage of 80 kV in the STEM experiment, which reduced the damage to the sample and gave satisfactory results. Therefore, the STEM characterization and analysis presented in Figure 2 are reliable.

## 3. Noise current and dark current

According to previous studies [Science 325,5948 (2009)], the total noise of photodetectors operating in DC and low frequency modes consists of two main components, thermal noise and shot noise. The shot noise originates from dark current. Their contributions can be estimated by calculations using the following parameters, $I_{dark}$ = 7 pA, $V_{ds}$ =5 V and $T$ = 300 K.

The thermal noise is calculated as

$$i_{thermal} = \sqrt{\frac{4k_B T}{R}} = \sqrt{\frac{4 \times 1.38 \times 10^{-23} \times 300}{7.14 \times 10^{11}}} \text{ A Hz}^{-1/2} = 1.52 \times 10^{-16} \text{ A Hz}^{-1/2},$$

where $k_B$ is the Boltzmann constant and $R$ is the channel resistance of the device.

The shot noise is calculated as

$$i_{shot} = \sqrt{2e I_{dark}} = \sqrt{2 \times 1.6 \times 10^{-19} \times 7 \times 10^{-12}} \text{ A Hz}^{-1/2} = 1.5 \times 10^{-15} \text{ A Hz}^{-1/2},$$

where $e$ is the elementary charge.

Thus, the total noise current is

$$i_{total} = \sqrt{i_{thermal}^2 + i_{shot}^2} = \sqrt{\left(1.52 \times 10^{-16}\right)^2 + \left(1.5 \times 10^{-15}\right)^2} \text{ A Hz}^{-1/2} = 1.507 \times 10^{-15} \text{ A Hz}^{-1/2} \approx i_{shot}.$$

From the calculations, it is found that the shot noise is one order of magnitude larger than the thermal noise. The shot noise is the major contribution to the total noise. Therefore, it is reasonable and safe to use $I_{dark}$ instead of the total noise current to calculate $D$*.